\documentclass[10pt,aps,prb,twocolumn]{revtex4-2}
\usepackage[english]{babel}
\usepackage[T1]{fontenc} %newley package
\usepackage[utf8]{inputenc} 
\usepackage{verbatim}
\usepackage{graphicx}
\usepackage{graphics}
\usepackage{color}
\usepackage{textcomp}
\usepackage{amsfonts}
\usepackage{amsmath}
\usepackage{hyperref}
\usepackage{amssymb}
\usepackage{mathtools}
\usepackage{tabularx}
\usepackage[a4paper, margin=1.5cm]{geometry}

\newcolumntype{Y}{>{\centering\arraybackslash}X}

\begin{document}

\title{Kinetic Flat-Histogram Simulations of Non-Equilibrium Stochastic Processes with Continuous and Discontinuous Phase Transitions}

% Authors: L. M. C. Alencar, Tayroni Francisco de Alencar Alves, Gladstone de Alencar Alves, Francisco Welington de Sousa Lima, Antônio Macedo Filho e Ronan Silva Ferreira

\author{L. M. C. Alencar}
\affiliation{Departamento de F\'{\i}sica, Universidade Federal do Piau\'{i}, 57072-970, Teresina - PI, Brazil}
\author{T. F. A. Alves}
\affiliation{Departamento de F\'{\i}sica, Universidade Federal do Piau\'{i}, 57072-970, Teresina - PI, Brazil}
\author{G. A. Alves}
\affiliation{Departamento de F\'{i}sica, Universidade Estadual do Piau\'{i}, 64002-150, Teresina - PI, Brazil}
\author{F. W. S. Lima}
\affiliation{Departamento de F\'{\i}sica, Universidade Federal do Piau\'{i}, 57072-970, Teresina - PI, Brazil}
\author{A. Macedo-Filho}
\affiliation{Departamento de F\'{i}sica, Universidade Estadual do Piau\'{i}, 64002-150, Teresina - PI, Brazil}
\author{R. S. Ferreira}
\affiliation{Departamento de Ci\^{e}ncias Exatas e Aplicadas, Universidade Federal de Ouro Preto, 35931-008, Jo\~{a}o Monlevade - MG, Brazil}

\date{Received: date / Revised version: date}

\begin{abstract}

To our knowledge, there is no flat-histogram algorithm to sample the stationary distribution of non-equilibrium stochastic processes. The present work addresses this gap by introducing a generalization of the Wang-Landau algorithm, applied to non-equilibrium stochastic processes with local transitions. The main idea is to sample macroscopic states using a kinetic Monte Carlo algorithm to generate trial moves, which are accepted or rejected with a probability that depends inversely on the stationary distribution. The stationary distribution is refined through the simulation by a modification factor. A visitation histogram is also accumulated, and the modification factor is updated when the histogram satisfies a flatness condition. An estimate of the stationary distribution is obtained in the limit where the modification factor reaches a threshold value close to unity. To test the algorithm, we compare simulation results for several stochastic processes with theoretically known behavior. In addition, results from the kinetic flat-histogram algorithm are compared with standard exact stochastic simulations. We show that the kinetic flat-histogram algorithm can be applied to phase transitions in stochastic processes with bistability, which describe a wide range of phenomena such as epidemic spreading, population growth, chemical reactions, and consensus formation. With some adaptations, the kinetic flat-histogram algorithm can also be applied to stochastic models on lattices and complex networks.

\end{abstract}



\maketitle

%%%%%%%%%%%%%%%%%%%%%%
\section{Introduction}
%%%%%%%%%%%%%%%%%%%%%%

The Wang-Landau algorithm~\cite{Wang-2001-1, Wang-2001-2, Landau-2004, Tsai-2008, Landau-2021} is a stochastic sampling method that enables the estimation of the density of states $g(E)$ in statistical equilibrium models. Here, $g(E)$ denotes the relative degeneracy of a particular energy state and represents its statistical weight in the microcanonical ensemble.

Whereas standard Monte Carlo methods, such as the Metropolis-Hastings algorithm, rely on importance sampling of the equilibrium canonical distribution at a fixed temperature~\cite{Landau-2021}, the Wang-Landau algorithm is designed to sample the energy space uniformly. It is also related to reweighting techniques, while offering the advantages of simplicity and straightforward implementation~\cite{Landau-2021}.

The Wang-Landau algorithm performs a random walk in the space of energy states $E_i$ of equilibrium models. It begins with an initial guess for the density of states, $g(E) = 1$, and updates it in a non-Markovian manner. Energy transitions are attempted with probability
\begin{equation}
   p(E_i \to E_f) = \min\left(1, \frac{g(E_i)}{g(E_f)}\right),
   \label{wl-prob}
\end{equation}
where $E_i$ and $E_f$ are the energy states before and after the trial, respectively. After each transition, the density of states is updated as
\begin{equation}
   \ln g(E) \to \ln g(E) + \ln f,
\end{equation}
where $f$ is a modification factor, initially set to a value greater than one, typically $f_0 = e \approx 2.71828$. If a transition is accepted, the density of states is updated at $E_f$; otherwise, it is updated at $E_i$.

Additionally, a histogram $H(E)$ is accumulated to count the number of visits to each energy state. The modification factor is reduced, for example, by $f \to \sqrt{f}$ whenever the histogram satisfies a flatness criterion. This criterion is typically defined such that the difference between any histogram bin and the average histogram value does not exceed a specified threshold. The algorithm terminates when the modification factor satisfies $1 < f < f_\text{final}$, where $f_\text{final}$ is a threshold value close to unity.

A common issue with the Wang-Landau method is error saturation~\cite{Belardinelli-2007-2}. One might expect that as the modification factor converges to unity, $g(E)$ would also converge. However, the deviation from the true density of states saturates, which can be mitigated by subtle modifications to the original algorithm~\cite{Belardinelli-2007-1, Belardinelli-2007-2, Zhou-2008}. Due to this error saturation, it is generally recommended to compare Wang-Landau results with standard Monte Carlo averages. Ongoing research continues to improve the Wang-Landau algorithm and expand its applications~\cite{Troyer-2003, Zhou-2006, Maerzke-2014, Vogel-2014, Chan-2017, Egorov-2018, Langfeld-2022}. Another example of a flat-histogram algorithm is the multicanonical Monte Carlo method~\cite{Liang-2007, Liang-2007-2, Liang-2009, Liang-2011}, which has been shown not to suffer from error saturation.

Additionally, it is important to note that the Wang-Landau algorithm does not satisfy the detailed balance condition during the initial stages of the random walk~\cite{Wang-2001-1, Landau-2004}. However, as the estimated density of states approaches the true density of states, detailed balance is restored for the stationary density of states. The Wang-Landau algorithm enables the direct calculation of the Helmholtz free energy and entropy in equilibrium systems, quantities that are not readily accessible through standard Monte Carlo methods sampling the canonical ensemble~\cite{Landau-2021}.

The Wang-Landau algorithm has been successfully applied to a wide range of equilibrium models, including spin systems~\cite{Silva-2006, Tsai-2008, Jorge-2019, Jorge-2021}, lattice gases~\cite{Jaleel-2021}, protein folding~\cite{Rathore-2004}, polymer models~\cite{Cunha-2006}, and quantum systems~\cite{Troyer-2003}. Furthermore, it can be used to study systems with continuous degrees of freedom~\cite{Zhou-2006, Sinha-2009} and complex energy landscapes~\cite{Wang-2001-2, Liang-2007, Liang-2007-2, Liang-2009, Liang-2011}. The Wang-Landau algorithm also enables the direct evaluation of the latent heat in bistable equilibrium systems.

However, to our knowledge, there is no flat-histogram algorithm for sampling the stationary distribution of non-equilibrium stochastic processes~\cite{Odor-2004, Hinrichsen-2006, Henkel-2008}, some of which exhibit bistable behavior~\cite{Schlogl-1972, Grassberger-1981, Grassberger-1982, Wang-2015, Ziff-1986, Voigt-1997, Oliveira-2018, Oliveira-2016, Fernandes-2018, Hinrichsen-2000, Vilela-2020}. This work addresses this gap by introducing a generalization of the Wang-Landau algorithm, applied to non-equilibrium stochastic processes with local transitions that obey the following master equation
\begin{eqnarray}
   \frac{d}{dt}\mathcal{P}(\boldsymbol{\sigma},t) = \sum_{n=1}^N &&
      \left[ w_n\!\left(\boldsymbol{\sigma}^{(n)}\to \boldsymbol{\sigma}\right)\,
      \mathcal{P}(\boldsymbol{\sigma}^{(n)},t) \right. \nonumber \\
   && - \left. w_n\!\left(\boldsymbol{\sigma}\to \boldsymbol{\sigma}^{(n)}\right)\,
      \mathcal{P}(\boldsymbol{\sigma},t) \right].
\label{masterequation}
\end{eqnarray}
for the case of one degree of freedom. Generalization to stochastic processes with more than one degree of freedom is straightforward. The master equation is a balance equation that gives $\mathcal{P}_{\boldsymbol{\sigma}}$, the probability distribution of the system states $\boldsymbol{\sigma}$ which are given as
\begin{equation}
   \boldsymbol{\sigma} = \left(\sigma_1, \sigma_2, \ldots, \sigma_n, \ldots, \sigma_N \right),
   \label{system-state}
\end{equation}
where each state is composed of $N$ local stochastic variables $\sigma_i$ ($i=1,2,\ldots,N$). Moreover, the master equation above describes Markovian dynamics with local transition rates $w_n(\boldsymbol{\sigma})$ among the system states. A local transition, i.e., a change of a local stochastic variable $\sigma_n \to \sigma^\prime_n$ with rate $w_n(\boldsymbol{\sigma})$, yields a new system state $\boldsymbol{\sigma}^{(n)}$ written as
\begin{equation}
   \boldsymbol{\sigma}^{(n)} = \left(\sigma_1, \sigma_2, \ldots, \sigma^\prime_n, \ldots, \sigma_N \right).
   \label{new-system-state}
\end{equation}

Our main idea is to implement a random walk in phase space to estimate the stationary distribution $P(\sigma)$ of the main system observable $\sigma \equiv \langle \boldsymbol{\sigma} \rangle$, which obeys the following time-evolution equation for one degree of freedom
\begin{equation}
   \frac{d\sigma}{dt} = -2 \left\langle \sigma_n w_n\!\left(\boldsymbol{\sigma}\to \boldsymbol{\sigma}^{(n)}\right)\right\rangle,
\end{equation}
obtained directly from Eq.~(\ref{masterequation}) for a homogeneous system, where $P(\sigma)$ is related to the stationary system distribution $\mathcal{P}(\boldsymbol{\sigma})$ by
\begin{equation}
   P(\sigma) = \sum_{\langle \boldsymbol{\sigma} \rangle = \sigma}^{\Omega} \mathcal{P}(\boldsymbol{\sigma}),
   \label{observable-evolution}
\end{equation}
with $\Omega$ being the total number of states. We then introduce weights in the random walk transitions to encourage visits to less frequently visited states, and the stationary distribution of the observable $\sigma$ is obtained from these weights when the visitation histogram becomes flat. Generalization to non-homogeneous systems is straightforward. We consider cases where the degrees of freedom are bounded and the dynamics can be entirely described by a single observable.

The concept of equilibrating transition rates is similar to that in Ref.~\cite{Lidmar-2012}. However, the kinetic flat-histogram algorithm does not sample the Gibbs free energy, nor do the rates depend on the energy. Additionally, we maintain a histogram of the observable $\sigma$. Some of the tested models are described solely by master equations and do not possess an associated Hamiltonian.

A potential application is the identification of discontinuous non-equilibrium phase transitions, i.e., non-equilibrium systems that exhibit bistability. We present simulation results for several well-known bistable systems. A bistable non-equilibrium system can display an exponential decay of the main observable at the threshold~\cite{Ptaszynski-2024}. However, the kinetic flat-histogram algorithm enables direct analysis of the stationary behavior.

The most distinctive feature of the kinetic flat-histogram algorithm is its extension of the Wang-Landau algorithm to systems not described by an energy landscape. In this approach, simulations are performed separately for each value of the control parameter (in equilibrium systems, this parameter is typically the temperature), which makes the kinetic flat-histogram algorithm more closely related to multicanonical Monte Carlo algorithms~\cite{Landau-2021}. Consequently, the stationary distribution obtained is $P(\sigma,\lambda)$, where $\lambda$ is the control parameter.

We organize this paper as follows. In Sec.~\ref{sec-model}, we present the kinetic flat-histogram algorithm. In Sec.~\ref{sec-results-continuous}, we show results for several mean-field stochastic processes with continuous phase transitions. In Sec.~\ref{sec-results-discontinuous}, we report findings for some bistable non-equilibrium stochastic processes. Finally, we provide our concluding remarks in Sec.~\ref{sec-conclusions}.

%%%%%%%%%%%%%%%%%%%%%%%%%%%%%%%%%%%%%%%%%%%%%%%%%%%%%%%%%%%
\section{Kinetic Flat-Histogram Algorithm}\label{sec-model}
%%%%%%%%%%%%%%%%%%%%%%%%%%%%%%%%%%%%%%%%%%%%%%%%%%%%%%%%%%%

In this section, we present the kinetic flat-histogram algorithm. In principle, the process should be Markovian with local updates, and the stochastic process should be ergodic. Some stochastic processes considered in this work do not fulfill this condition because they display absorbing states. The presence of absorbing states can lead to non-equilibrium phase transitions, with the most common universality class being directed percolation~\cite{Odor-2004, Hinrichsen-2006, Henkel-2008}. Additionally, the stochastic process should obey the Perron-Frobenius theorem~\cite{vanKampen-1992, Tome-2015}, ensuring a well-defined stationary distribution.

Absorbing states can be addressed by perturbing the non-equilibrium dynamics with reactivation~\cite{Macedo-2018}, which prevents the dynamics from being trapped. This approach ensures ergodicity and allows the simulation to continue. The main effect is that the stationary distribution is no longer trivial, since any dynamics would otherwise be trapped in the absorbing state for finite system sizes. Spontaneous particle insertion when the system visits the absorbing state is equivalent to an external field that scales as $1/N$ away from the threshold (in fact, the external field scales as $N^{-\sigma}$ at the threshold).

We consider well-studied stochastic processes in the mean-field regime, where exact results exist. By hypothesis, we can generate local transitions, referred to as \textbf{trials}, that update the main observable $\sigma_i \to \sigma_f$. Trials can be performed using standard Monte Carlo algorithms. For mean-field (one-site) stochastic processes, the exact stochastic simulation algorithm (SSA)~\cite{Gillespie-1976, Gillespie-1977} is used. For lattice models and stochastic processes on graphs and networks~\cite{Newman-2018}, the kinetic Monte Carlo algorithm~\cite{Landau-2021} can be employed.

In the following sections, we discuss several reactive processes as prototypes of non-equilibrium stochastic systems with phase transitions. For one-site reactive processes with a single degree of freedom~\cite{Tome-2015}, described by the reversible reaction
\begin{equation}
   \sum_{\ell=1}^q \nu_\ell^{\prime} A_\ell \xrightleftharpoons[k^-]{k^+}  \sum_{\ell=1}^q \nu_\ell^{\prime\prime} A_\ell,
\end{equation}
where $A_\ell$ are the $q$ chemical species and $\nu_\ell^{\prime}$, $\nu_\ell^{\prime\prime}$ are the respective stoichiometric coefficients, Eq.~(\ref{masterequation}) can be rewritten as
\begin{eqnarray}
   \frac{d}{dt} P_\sigma(t) &=& a_{\sigma-1} P_{\sigma-1}(t) + b_{\sigma+1} P_{\sigma+1}(t) \nonumber \\
                            & & - \left( a_\sigma + b_\sigma \right)  P_\sigma(t),
   \label{local-master-equation}
\end{eqnarray}
where $a_\sigma$ and $b_\sigma$ are particular cases of the general transition rates $w_n$. Eq.~(\ref{local-master-equation}) describes a standard birth-death process~\cite{Tome-2015}.

From Eq.~(\ref{local-master-equation}), in the continuous limit, we can obtain the following Langevin equation for the main observable
\begin{equation}
   \frac{d \sigma}{dt} = \left\langle a(\sigma) - b(\sigma) \right\rangle
                       + \frac{2}{N} \left\langle a(\sigma) + b(\sigma) \right\rangle \zeta(\sigma,t),
   \label{langevin-reactive}
\end{equation}
where $\left\langle \zeta(\sigma,t) \right\rangle = 0$ and $\left\langle \zeta(\sigma,t) \zeta(\sigma,t^\prime) \right\rangle = \delta(t-t^\prime)$. For one-site reactive processes, $a(\sigma)$ and $b(\sigma)$ in Eq.~(\ref{langevin-reactive}) can be calculated as~\cite{Tome-2015}
\begin{equation}
      a(\sigma) = k^+ \prod_{\ell=1}^{q} \left[x_\ell(\sigma)\right]^{\nu^{\prime}_\ell}, \quad
      b(\sigma) = k^- \prod_{\ell=1}^{q} \left[x_\ell(\sigma)\right]^{\nu^{\prime\prime}_\ell},
   \label{reaction-extensions}
\end{equation}
where $x_\ell = N_\ell/N$ are the concentrations of the chemical species $A_\ell$, given as functions of the observable $\sigma$. The time-evolution equation for the main observable $\sigma$ can be obtained by averaging the Langevin equation. Thus, the stochastic process admits two equivalent descriptions: the master equation and the Langevin equation, both of which can be addressed using the kinetic flat-histogram algorithm.

Generalization to more than one degree of freedom (i.e., additional reactions) and external sources is straightforward~\cite{Tome-2015}. The result presented in Eq.~(\ref{langevin-reactive}), and their generalized forms for multiple degrees of freedom are applied to the one-site stochastic processes discussed in Secs.~\ref{sec-results-continuous} and \ref{sec-results-discontinuous} to obtain the corresponding time-evolution equations.

In general, the transition rates are functions of an external control parameter $\lambda$ that drives a phase transition, where the order parameter is the average of the main observable. For each value of $\lambda$, we perform a distinct simulation that yields a unique stationary distribution $P(\sigma,\lambda)$, which is analogous to simulate the original Wang-Landau algorithm for various values of the magnetic field or the exchange coupling constant in a magnetic system. Given the trials $\sigma_i \to \sigma_f$, the essence of the kinetic flat-histogram algorithm is as follows:
\begin{enumerate}
   \item Initialize the system in a random state. Set the modification factor $f$ with $\ln f_0 = 1$. Also, initialize two arrays: $\ln P(\sigma,\lambda) = 0$ and $H(\sigma,\lambda) = 0$, which store the logarithm of the stationary distribution and the normalized visitation histogram for all possible values of the main observable $\sigma$, respectively. The arrays may be grown dynamically; in this case, a third array containing the visited $\sigma$ bins is added. Arrays should be resized and reordered when a new $\sigma$ is visited. At the end of the simulation, $P(\sigma,\lambda)$ will converge to the asymptotic stationary distribution. Comparing expected values from the stationary distribution with time-series averages from standard methods is recommended to test consistency;
   \item Begin updating the main observable. Accept trials with the probability
   \begin{equation}
      p(\sigma_i \to \sigma_f) = \frac{P(\sigma_i,\lambda)}{P(\sigma_i,\lambda)+P(\sigma_f,\lambda)},
      \label{flat-hist-prob}
   \end{equation}
   and update the arrays containing the stationary distribution and the visitation histogram as
   \begin{equation}
      \left\lbrace
      \begin{aligned}
         & \ln P(\sigma,\lambda) \to \ln P(\sigma,\lambda) + \ln f, \\
         & H(\sigma,\lambda) \to H(\sigma,\lambda) + 1,
      \end{aligned}
      \right.
   \end{equation}
   respectively, where $\sigma = \sigma_f$ if the trial is accepted, and $\sigma = \sigma_i$ if rejected. Also update the system state $\boldsymbol{\sigma} \to \boldsymbol{\sigma}^{(n)}$ and the value of the main observable $\sigma_i \to \sigma_f$ for accepted trials. A Monte Carlo step is defined as $N$ local updates;
   \item After a number of Monte Carlo steps (we used $10^4$ steps for all results), test the flatness condition for the histogram. The flatness condition requires that all histogram bins satisfy $0.95 \langle H \rangle \leq H(E_i) \leq 1.05 \langle H \rangle$, where $\langle H \rangle$ is the histogram average. If the flatness condition is satisfied, update the modification factor as
   \begin{equation}
      \ln f \to \ln f / \sqrt{N \ln 2};
      \label{factor-update}
   \end{equation}
   \item End the simulation when $0 < \ln f < \ln f_\text{final}$. In most results, we chose $\ln f_\text{final} = 10^{-7}$.
\end{enumerate}

After the simulation, we can calculate the expected value $\varsigma(\lambda)$ of the main observable $\sigma$ from the converged stationary distribution $P(\sigma, \lambda)$ as
\begin{equation}
   \varsigma(\lambda) = \sum_{\sigma} \sigma P(\sigma, \lambda).
   \label{order-parameter}
\end{equation}
Additionally, the kinetic flat-histogram algorithm enables the direct calculation of the Shannon entropy, defined as
\begin{equation}
   S(\lambda) = - \sum_{\sigma} P(\sigma, \lambda) \ln P(\sigma, \lambda).
   \label{shannon-entropy}
\end{equation}
For equilibrium systems that obey detailed balance, the Shannon entropy coincides with the Boltzmann entropy.

In the following sections, we present results for the algorithm applied to several well-known stochastic processes.

%%%%%%%%%%%%%%%%%%%%%%%%%%%%%%%%%%%%%%%%%%%%%%%%%%%%%%%%%%%%%%%%%%%%
\section{Continuous Phase Transitions}\label{sec-results-continuous}
%%%%%%%%%%%%%%%%%%%%%%%%%%%%%%%%%%%%%%%%%%%%%%%%%%%%%%%%%%%%%%%%%%%%

We begin by applying the kinetic flat-histogram algorithm to several stochastic processes that exhibit continuous phase transitions. For all processes, we set the system size to $N=10^3$. The models considered are the Glauber model~\cite{Oliveira-2003, Janke-2019, Liu-2023}, the majority-vote (MV) model~\cite{Oliveira-1992, Pereira-2005, Wu-2010, Yu-2017, Crochik-2005, Vilela-2009, Lima-2012, Vieira-2016, Fronczak-2017, Lima-2006, Lima-2007, Vilela-2018, Krawiecki-2018, Krawiecki-2018-2, Alves-2019, Krawiecki-2019, Vilela-2020-2, Zubillaga-2022, Alencar-2023-1, Alencar-2024}, the contact process (CP)~\cite{Harris-1974, Jensen-1992, Hinrichsen-2000, Dickman-2002, Marro-2005, Castellano-2006, Hinrichsen-2006, Castellano-2008, Oliveira-2008, Henkel-2008, Boguna-2009, Munoz-2010, Ferreira-2011-1, Ferreira-2011-2, Odor-2012, Silva-2013, Ferreira-2013, Ferreira-2016, Almeida-2016, Bottcher-2018, Santos-2018, Alencar-2023-2}, and the first Schlögl model~\cite{Schlogl-1972, Grassberger-1981}.

%~~~~~~~~~~~~~~~~~~~~~~~~~
\subsection{Glauber Model}
%~~~~~~~~~~~~~~~~~~~~~~~~~

The Glauber model is a stochastic dynamics for the Ising model~\cite{Oliveira-2003, Janke-2019, Liu-2023}. Notably, the Glauber model obeys detailed balance, so in the asymptotic limit $t \to \infty$. Indeed, the Glauber dynamics is a relaxation dynamics for the equilibrium Ising model. In the mean-field (one-site) limit, the spin-flip reaction channels can be written as
\begin{equation}
   \downarrow \xrightarrow{k_1} \uparrow, \quad \text{and} \quad \uparrow \xrightarrow{k_2} \downarrow,
   \label{spin-flip}
\end{equation}
where $N = N_\uparrow + N_\downarrow$ is the system size. In Eq.~(\ref{spin-flip}), $k_1$ and $k_2$ are the spin-flip propensities
\begin{equation}
   \begin{aligned}
      & k_1 = \frac{1}{2}\left[1+\tanh\left(\lambda M\right)\right], \\
      & k_2 = \frac{1}{2}\left[1-\tanh\left(\lambda M\right)\right],
   \end{aligned}
   \label{propensities-glauber}
\end{equation}
where the main observable is the magnetization, given by $M = (N_\uparrow - N_\downarrow)/N$. The SSA simulation is straightforward: one of the two channels in Eq.~(\ref{spin-flip}) is chosen with a probability proportional to its propensity in Eq.~(\ref{propensities-glauber}).

The magnetization $M$ obeys the following time-evolution equation
\begin{equation}
   \frac{dM}{dt} = -M + \tanh\left(\lambda M\right),
\end{equation}
which predicts a continuous phase transition from a paramagnetic phase to a ferromagnetic phase, with an exact threshold at $\lambda_c = 1$. Note that the stationary magnetization satisfies the Bragg-Williams equation of state. The order parameter is the absolute value of the magnetization $m$, and the control parameter $\lambda$ acts as the inverse temperature.

\begin{figure}[tbp]
   \begin{center}
      \includegraphics[scale=0.285]{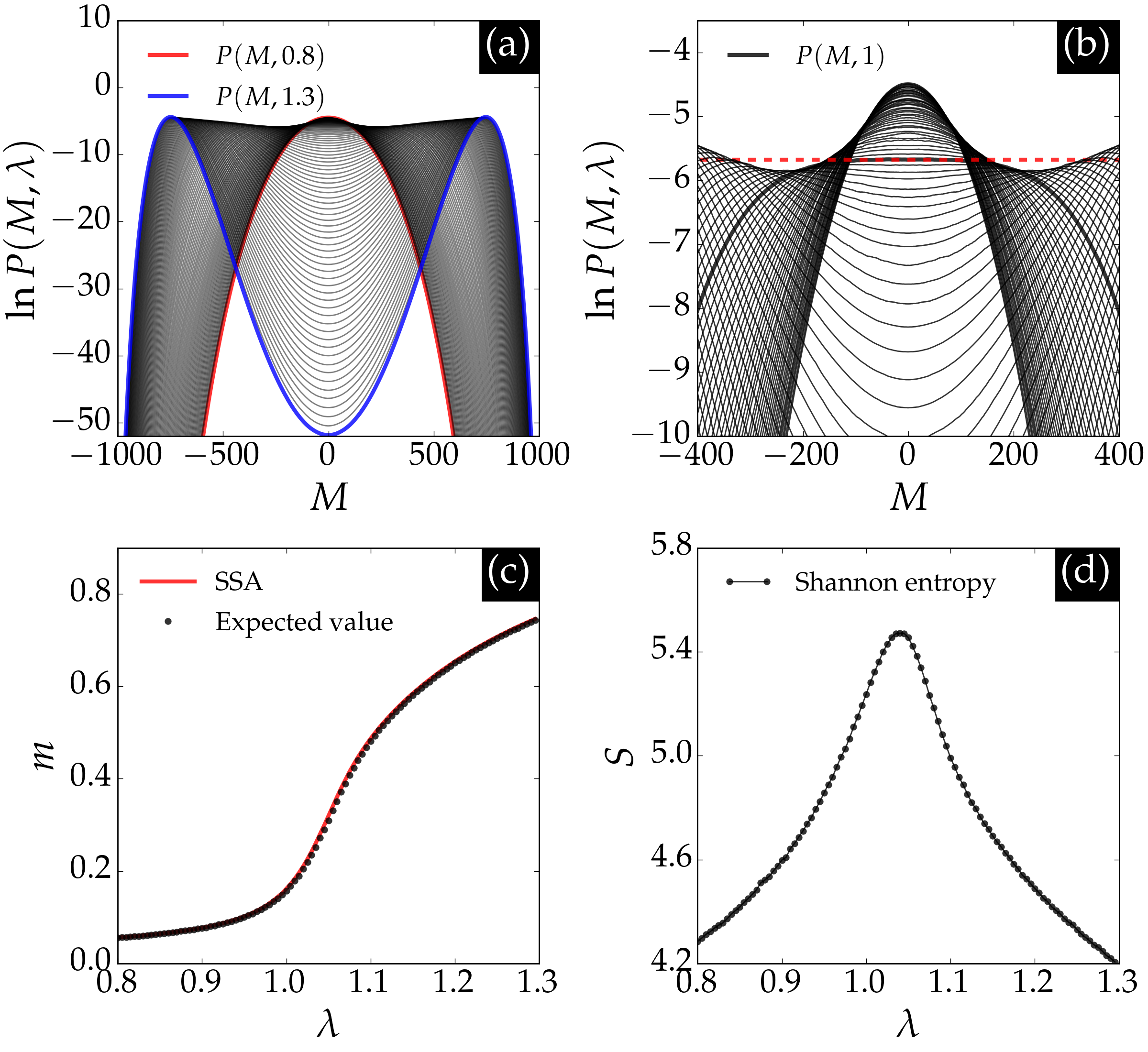}
   \end{center}
   \caption{(Color Online) Results for the Glauber model. Panel (a) shows the stationary distribution $P(M,\lambda)$ for a sequence of inverse temperatures $\lambda$. $P(M,\lambda)$ exhibits the expected behavior of a continuous phase transition, characterized by two symmetric maxima in the ferromagnetic phase ($\lambda>1$) and a single maximum in the paramagnetic phase ($\lambda<1$). Panel (b) presents a magnified view of panel (a), highlighting the critical $P(M,\lambda_c)$. The critical threshold $\lambda_c=1$ separates the two phases. Panel (c) displays the absolute value of the magnetization $m$, demonstrating good agreement between the expected value from $P(M,\lambda)$ and the time-series average from SSA simulations. Finally, panel (d) shows the Shannon entropy $S(\lambda)$, which is continuous as expected for a continuous phase transition. The line in panel (d) is only a guide to the eye.}
   \label{results-glauber}
\end{figure}

We present results for the kinetic flat-histogram algorithm in Fig.~(\ref{results-glauber}). The algorithm determines the stationary distribution $P(M,\lambda)$, which exhibits the expected behavior for an inversion-symmetry breaking transition. Specifically, the critical distribution $P(M,\lambda_c)$ separates the paramagnetic phase, where the stationary distribution has a single maximum at $M=0$, from the ferromagnetic phase, where it has two symmetric maxima. There is good agreement between the order parameter values obtained using the kinetic flat-histogram simulations and those from the SSA algorithm. Additionally, the continuity of the Shannon entropy is consistent with a continuous phase transition.

\begin{figure}[tbp]
   \begin{center}
      \includegraphics[scale=0.285]{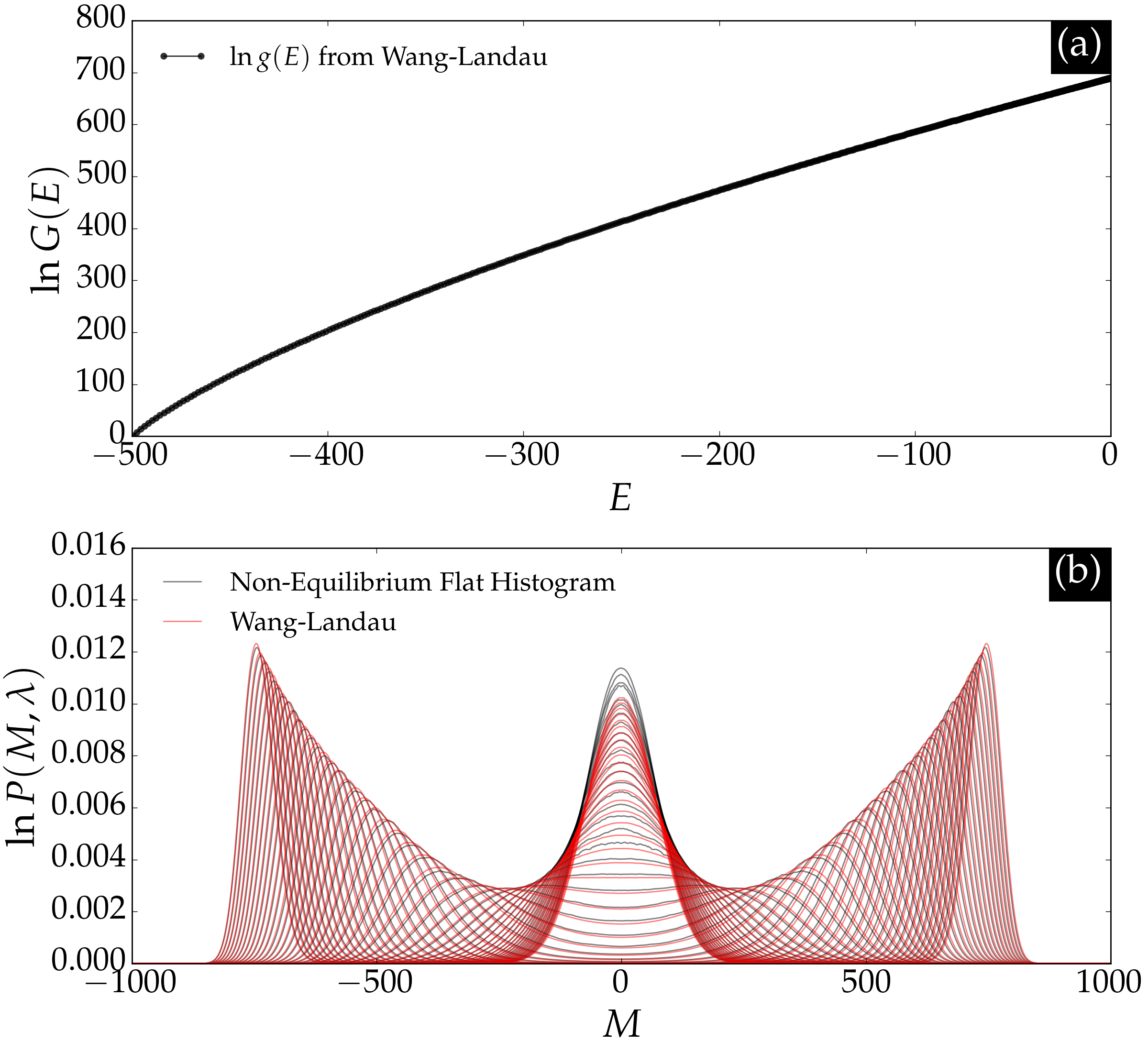}
   \end{center}
   \caption{(Color Online) Results for the Glauber model. Panel (a) shows the density of states $g(E)$ obtained using the original Wang-Landau algorithm. Panel (b) presents the stationary distribution $P(M,\lambda)$ calculated with both the kinetic flat-histogram algorithm and the original Wang-Landau algorithm, for 50 inverse temperatures $\lambda$ ranging from $0.8$ to $1.3$. The results are consistent, confirming the validity of the kinetic flat-histogram algorithm for equilibrium systems. However, the agreement deteriorates when going deep into the paramagnetic phase.}
   \label{FHvsWL}
\end{figure}

The Glauber model also allows us to test the kinetic flat-histogram algorithm for equilibrium systems, since the Glauber dynamics simulates the Ising model, whose Hamiltonian is given by
\begin{equation}
   \mathcal{H} = -\frac{J}{2N} \sum_{i,j} s_i s_j,
\end{equation}
where $s_i = \pm 1$ are Ising spins and $J=1$ is the exchange coupling constant. From the density of states $g(E)$, one can calculate the stationary distribution $P(E,\lambda)$ in the canonical ensemble using
\begin{equation}
   P(E,\lambda) = \frac{g(E) e^{-\lambda E}}{Z(\lambda)},
\end{equation}
where 
\begin{equation}
   E = -\frac{J}{2N} M^2
\end{equation}
is the mean-field energy and $Z(\lambda)$ is the partition function. We performed simulations using the original Wang-Landau algorithm to obtain $g(E)$ and calculated $P(M,\lambda)$ from it. The results agree with those obtained from the kinetic flat-histogram algorithm, confirming its validity for equilibrium systems, as shown in Fig.~(\ref{FHvsWL}). The data from both simulations are consistent. However, the agreement deteriorates when going deep into the paramagnetic phase.

%~~~~~~~~~~~~~~~~~~~~~~~~~~~~~~~
\subsection{Majority-Vote Model}
%~~~~~~~~~~~~~~~~~~~~~~~~~~~~~~~

We now consider the MV model~\cite{Oliveira-1992, Pereira-2005, Wu-2010, Yu-2017, Crochik-2005, Vilela-2009, Lima-2012, Vieira-2016, Fronczak-2017, Lima-2006, Lima-2007, Vilela-2018, Krawiecki-2018, Krawiecki-2018-2, Alves-2019, Krawiecki-2019, Vilela-2020-2, Zubillaga-2022, Alencar-2023-1, Alencar-2024}, a well-studied model of consensus formation. The MV model also describes a ferromagnetic material in contact with two heat reservoirs, one at $T=0$ and the other at $T\to\infty$. In the mean-field (one-site) regime, the model is equivalent to the Ehrenfest urn model.

On lattices and networks, the MV model is irreversible and displays a non-equilibrium phase transition. The MV transition in lattices and networks exhibits the same stationary exponents as the Ising model and belongs to the same universality class. In the mean-field (one-site) limit, the MV model obeys the same reaction channels as the Glauber model, given in Eq.~(\ref{spin-flip}). However, the propensities are given by
\begin{equation}
   \begin{aligned}
      & k_1 = \frac{1}{2}\left[1 + (1-2\lambda)\text{Sign}(M)\right], \\
      & k_2 = \frac{1}{2}\left[1 - (1-2\lambda)\text{Sign}(M)\right],
   \end{aligned}
   \label{propensities-mv}
\end{equation}
where the magnetization is still given by $M = (N_\uparrow - N_\downarrow)/N$. The propensities indicate that the spin opposes the majority direction with probability $\lambda$, follows the majority with probability $1-\lambda$, and has a random orientation if there is no majority.

The magnetization $M$ in the MV model obeys the following time-evolution equation
\begin{equation}
   \frac{dM}{dt} = -M + \left(1-2\lambda\right)\text{Sign}\left(M\right),
\end{equation}
where $\text{Sign}(x)$ is defined as
\begin{equation}
   \text{Sign}(x) = 
   \begin{cases}
      -1 & \text{if } x < 0, \\
       0 & \text{if } x = 0, \\
       1 & \text{if } x > 0, \\
   \end{cases}
\end{equation}
This equation predicts a continuous phase transition from a ferromagnetic phase to a paramagnetic one, with an exact threshold at $\lambda_c = 0.5$. The order parameter is the magnetization $M$, and the control parameter $\lambda$ acts as noise, defined as the probability that an individual opposes the opinion state of the majority.

\begin{figure}[tbp]
   \begin{center}
      \includegraphics[scale=0.285]{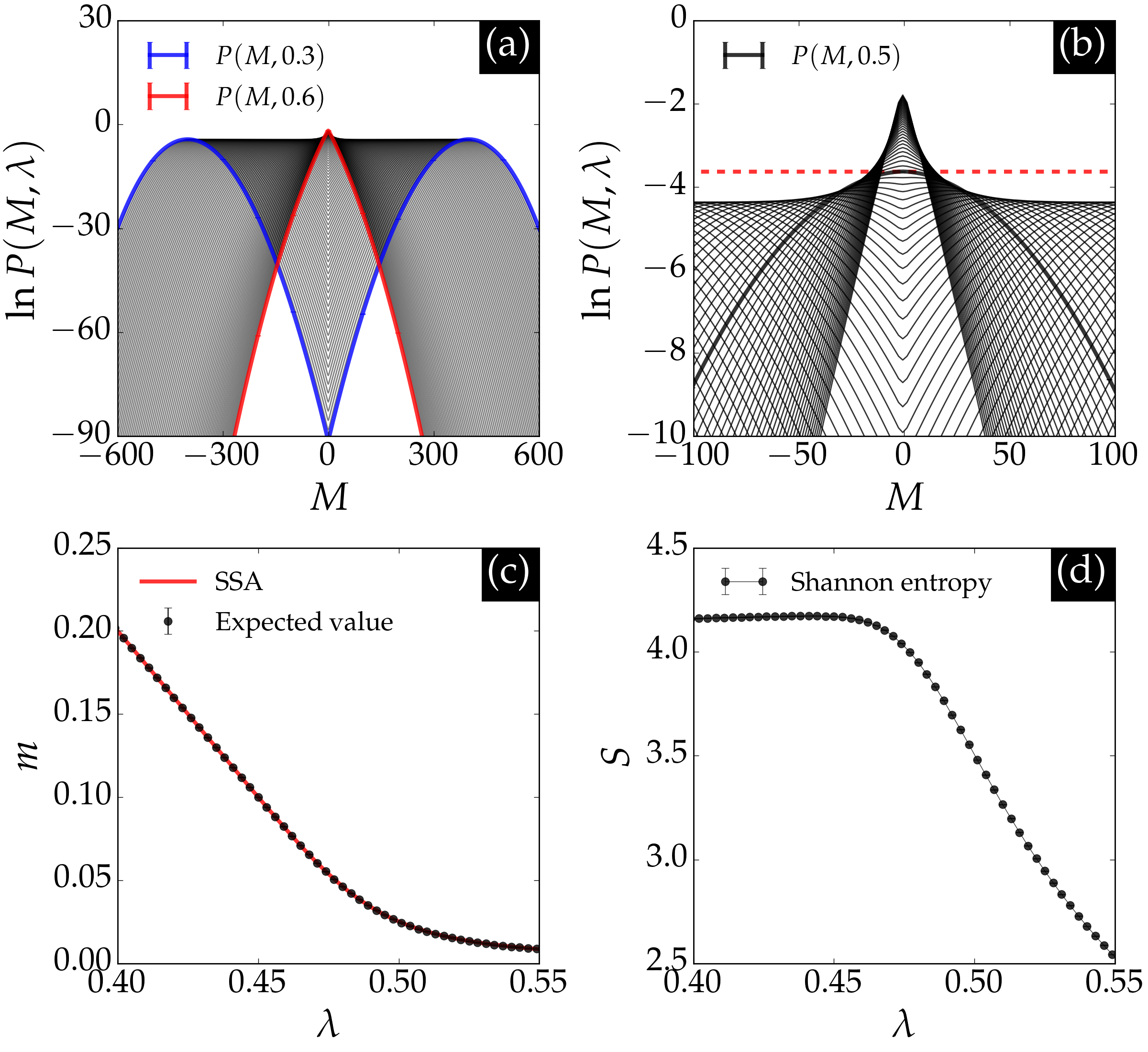}
   \end{center}
   \caption{(Color Online) Results for the MV model. As in Fig.~(\ref{results-glauber}), the system transitions from the ferromagnetic phase to the paramagnetic phase as $\lambda$ increases, with a threshold at $\lambda_c=0.5$. For this case, we also present statistical error bars for the order parameter $m$ obtained from the kinetic flat-histogram simulations. The error bars, calculated from 10 independent runs, are smaller than the symbol size. The line in panel (d) is shown only as a guide to the eye.}
   \label{results-mv}
\end{figure}

We present results for the kinetic flat-histogram algorithm in Fig.~(\ref{results-mv}). As expected for an inversion-symmetry breaking transition, the critical distribution $P(M,\lambda_c)$ separates the paramagnetic phase (characterized by a single maximum at $M=0$) from the ferromagnetic phase (with two symmetric maxima). The order parameter values obtained from both SSA and flat-histogram simulations are in good agreement. Furthermore, the Shannon entropy remains continuous, consistent with a continuous phase transition. In Fig.~(\ref{results-mv}), we also show statistical error bars for the order parameter $m$ obtained from 10 independent runs of the kinetic flat-histogram algorithm; these error bars are smaller than the symbol size.

\begin{figure}[tbp]
   \begin{center}
      \includegraphics[scale=0.285]{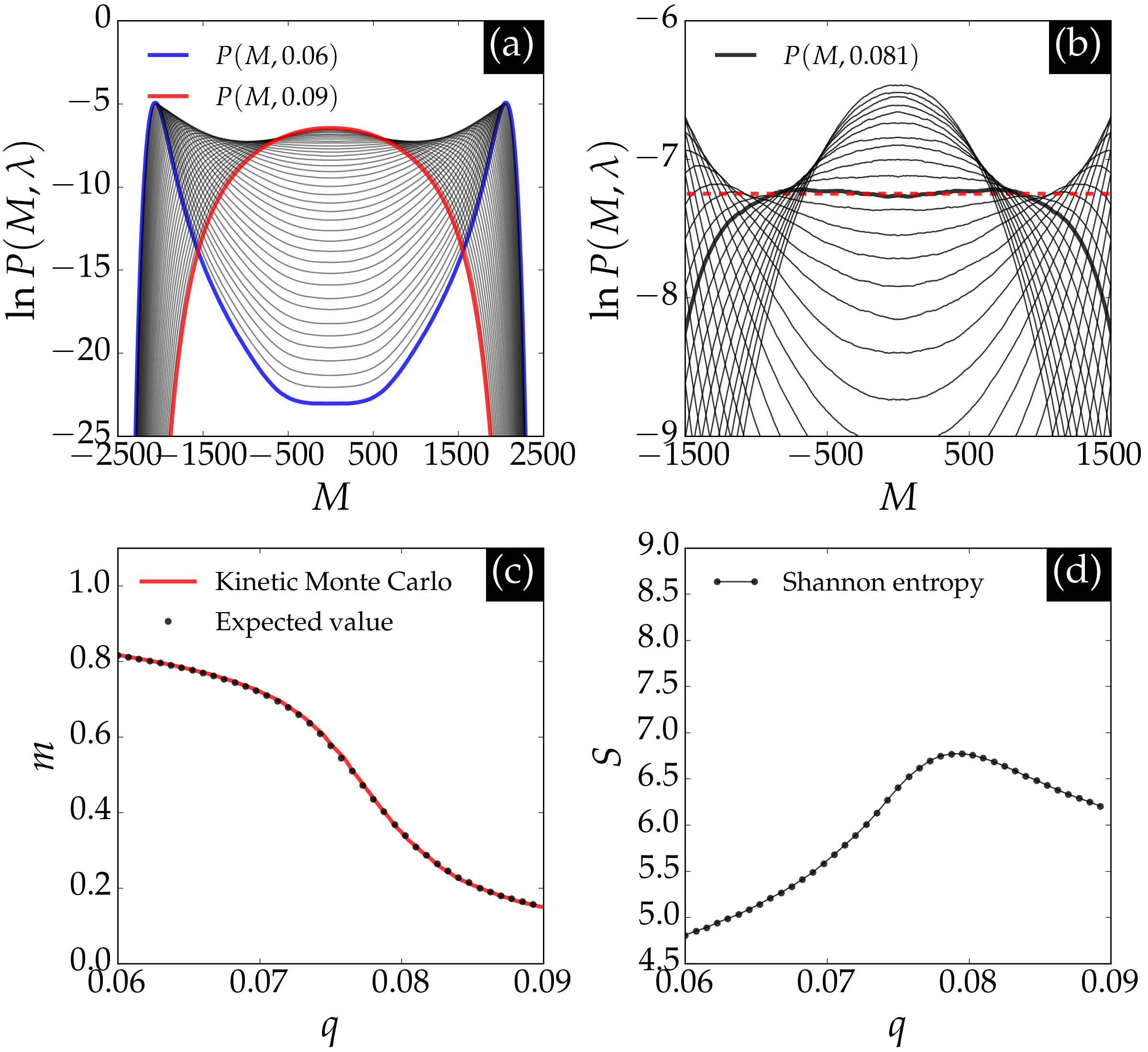}
   \end{center}
   \caption{(Color Online) Results for the MV model on a square lattice of size $N=50\times50$. As in Fig.~(\ref{results-glauber}), the transition from the two-maxima regime to the single-maximum regime allows us to identify a pseudo-critical point $q^\prime = 0.081$, which is close to the maximum of the Shannon entropy. The line in panel (d) is shown only as a guide to the eye.}
   \label{results-mv-lattice}
\end{figure}

\begin{figure}[tbp]
   \begin{center}
      \includegraphics[scale=0.285]{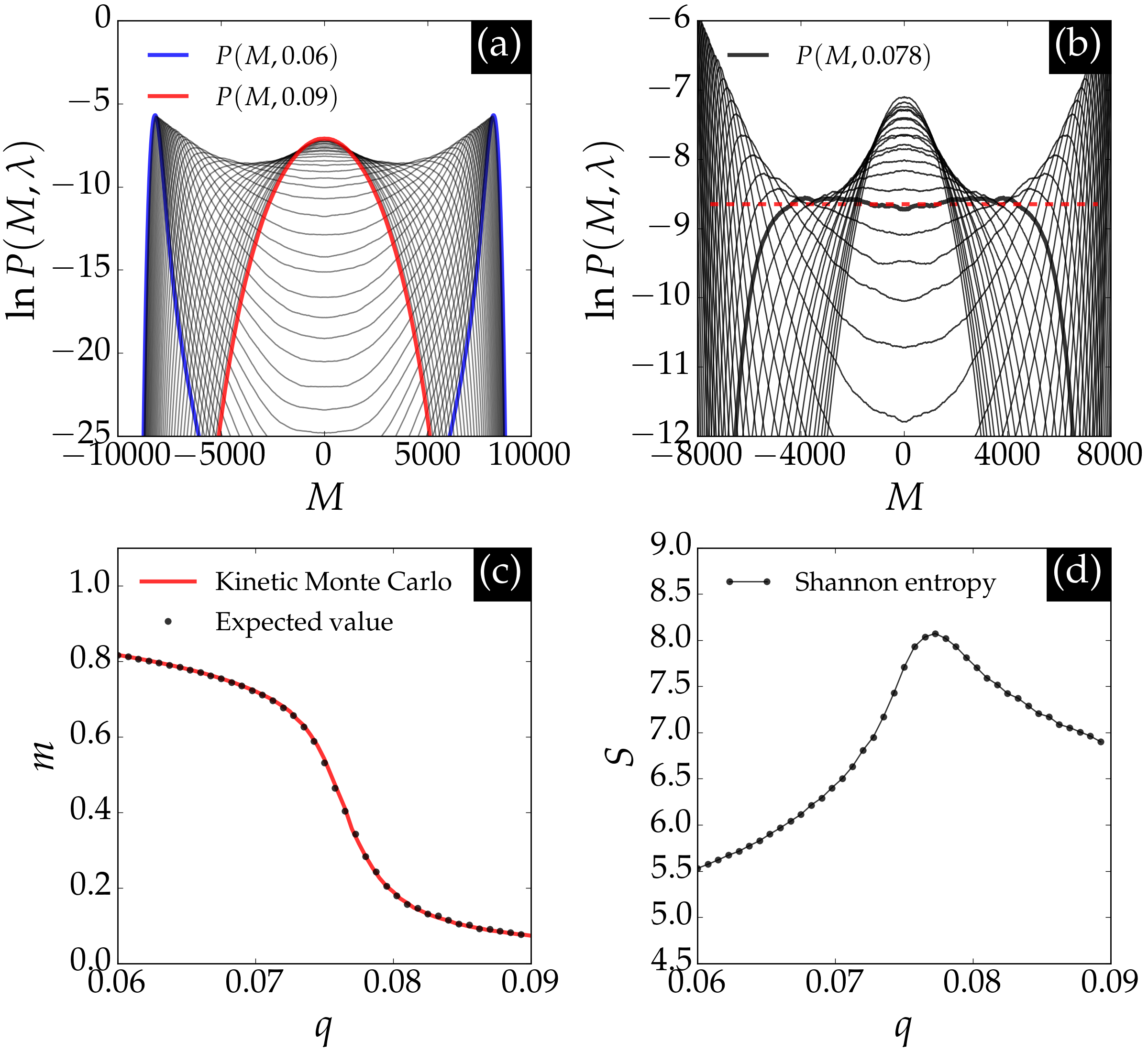}
   \end{center}
   \caption{(Color Online) Same as Fig.~(\ref{results-mv-lattice}), except for $N=100\times100$. We note that the pseudo-critical point $q^\prime = 0.078$ approaches the critical threshold $q_c=0.075$ as the lattice size increases.}
   \label{results-mv-lattice-2}
\end{figure}

To demonstrate the applicability of the method also to lattice systems, we present results for the MV model on square lattices in Fig.~(\ref{results-mv-lattice}) and~(\ref{results-mv-lattice-2}). New system states are generated using the kinetic Monte Carlo algorithm~\cite{Landau-2021}, where spin flips are attempted on randomly chosen nodes with rate
\begin{equation}
   w(\sigma_{i}) = \frac{1}{2}\left[1-\left(1-2q\right)\sigma_{i}\text{Sign}\left(\sum_{\left<i,j\right>} \sigma_{j} \right)\right],
\end{equation}
with the summation performed over nearest neighbors and the sign function as previously defined. The local stochastic variable $\sigma_i = \pm 1$ represents the spin state of node $i$, and $q$ is the noise parameter. The kinetic Monte Carlo algorithm is equivalent to the SSA for lattice systems, and the time unit is defined as $N$ trials, where $N=L\times L$ is the system size.

The spin-flip rate corresponds to the spin following the majority opinion in its neighborhood with probability $1-q$, and the opposite with probability $q$, where $q$ acts as a social temperature by introducing local contrarians. After generating a trial move via kinetic Monte Carlo, the kinetic flat-histogram algorithm is applied by accepting trials with the probability given in Eq.~(\ref{flat-hist-prob}). Alternatively, one may use the product of the spin-flip probability and the flat-histogram acceptance probability to generate system states. The results from kinetic Monte Carlo and flat-histogram simulations are in good agreement, as shown in Figs.~(\ref{results-mv-lattice}) and~(\ref{results-mv-lattice-2}).

The results for square lattices are qualitatively similar to those in the mean-field regime, except for finite-size effects. The transition from two maxima to one in the stationary distribution enables the detection of pseudo-critical points $q^\prime$ at noise values $q=0.081$ and $0.078$ for $L=50$ and $L=100$, respectively. The pseudo-critical point corresponds to the maximum of both the magnetic susceptibility and the Shannon entropy, and approaches the true critical point as the system size increases, following the scaling relation $q^\prime - q_c \sim L^{-1/\nu}$~\cite{Lima-2012, Vilela-2020-2, Zubillaga-2022}. Additionally, the continuous Shannon entropy is consistent with a continuous phase transition.

%~~~~~~~~~~~~~~~~~~~~~~~~~~~
\subsection{Contact Process}
%~~~~~~~~~~~~~~~~~~~~~~~~~~~

The contact process (CP)~\cite{Harris-1974, Jensen-1992, Hinrichsen-2000, Dickman-2002, Marro-2005, Castellano-2006, Hinrichsen-2006, Castellano-2008, Oliveira-2008, Henkel-2008, Boguna-2009, Munoz-2010, Ferreira-2011-1, Ferreira-2011-2, Odor-2012, Silva-2013, Ferreira-2013, Ferreira-2016, Almeida-2016, Bottcher-2018, Santos-2018, Alencar-2023-2} is a well-studied stochastic process belonging to the directed percolation universality class. It is the simplest stochastic process that exhibits an absorbing state. In the mean-field (one-site) regime, the reaction channels for the CP are given by
\begin{equation}
   S + I \xrightarrow{k_1} 2I, \quad \text{and} \quad I \xrightarrow{k_2} S,
   \label{channels-cp}
\end{equation}
where $S$ denotes a susceptible individual and $I$ denotes an infected one. The first channel in Eq.~(\ref{channels-cp}) is an autocatalytic reaction representing contamination, with propensity $k_1 = \lambda$. The second channel describes spontaneous recovery, with propensity $k_2 = 1$. The dynamic behavior of the model is characterized by the infected concentration $\rho = N_I/N$, where the system size is $N = N_S + N_I$. Note that $\rho=0$ is an absorbing state.

The infected concentration evolves according to the following time-evolution equation
\begin{equation}
   \frac{d\rho}{dt} = \lambda\rho\left(1 - \rho\right) - \rho,
\end{equation}
which predicts an active-absorbing phase transition with a threshold at $\lambda_c=1$, where $\rho$ serves as the order parameter. The CP can be used to describe epidemic spreading without permanent immunity, with $\lambda$ representing the contamination probability and the recovery probability set to unity. For $\lambda<1$, the system is in the absorbing phase with $\rho=0$, while for $\lambda>1$, it is in the active phase with $\rho \ne 0$.

\begin{figure}[tbp]
   \begin{center}
      \includegraphics[scale=0.285]{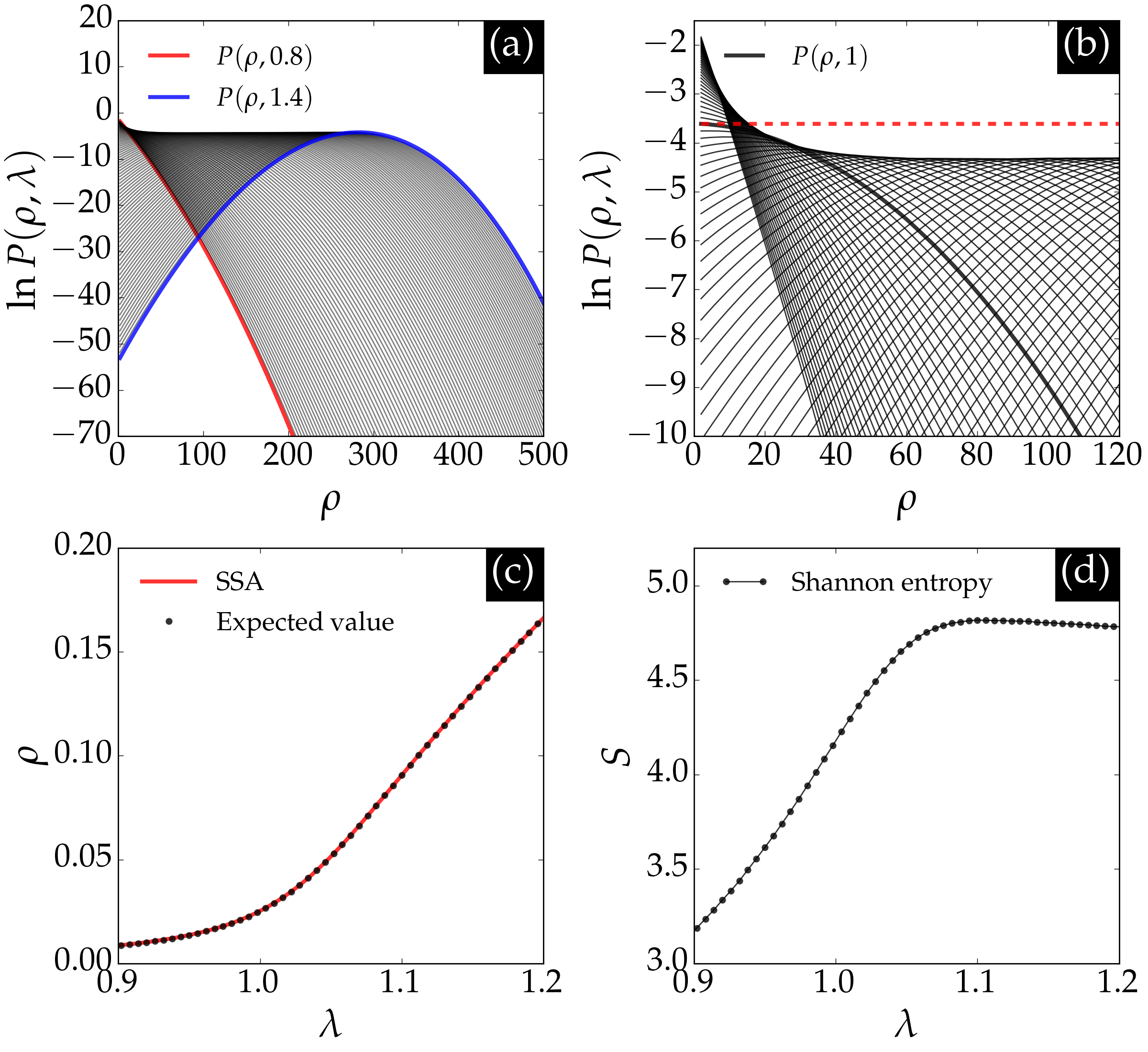}
   \end{center}
   \caption{(Color Online) Results for the CP model. Panel (a) shows the stationary distribution $P(\rho,\lambda)$ for a sequence of contamination probabilities $\lambda$. For simplicity, $P(1,\lambda)$ and $P(2,\lambda)$ are not shown, as these points are affected by the reactivation method~\cite{Macedo-2018}. The reactivation method is necessary to ensure ergodicity in the CP dynamics, which is required for kinetic flat-histogram simulations. In the absorbing phase, $P(\rho,\lambda)$ exhibits a single maximum at the absorbing state $\rho=0$. Panel (b) presents a magnified view, highlighting the critical $P(\rho,\lambda_c)$ for the CP. The critical $P(\rho,\lambda_c)$ approaches the absorbing state $\rho=0$ with a vanishing derivative. Panel (c) demonstrates good agreement between the expected value from Eq.~(\ref{order-parameter}) and the time-series average from the SSA algorithm. Finally, panel (d) shows the Shannon entropy $S(\lambda)$, which is continuous as expected for a continuous phase transition. The line in panel (d) is shown only as a guide to the eye.}
   \label{results-cp}
\end{figure}

We present results for the kinetic flat-histogram algorithm in Fig.~(\ref{results-cp}). The absorbing-active phase transition is characterized by a positive-definite order parameter. The critical distribution $P(\rho,\lambda_c)$ separates the absorbing phase ($\lambda < 1$) from the active phase ($\lambda > 1$). In the absorbing phase, the stationary distribution $P(\rho,\lambda)$ exhibits a single maximum at the absorbing state $\rho=0$. In the active phase, $P(\rho,\lambda)$ displays a local minimum at the absorbing state. At the threshold, $P(\rho,\lambda)$ approaches the absorbing state with a vanishing derivative. Additionally, the continuity of the Shannon entropy is consistent with a continuous phase transition.

%~~~~~~~~~~~~~~~~~~~~~~~~~~~~~~~
\subsection{First Schlögl Model}
%~~~~~~~~~~~~~~~~~~~~~~~~~~~~~~~

The first Schlögl model~\cite{Schlogl-1972, Grassberger-1981} was introduced as a prototype of an open chemical reactive system exhibiting a continuous phase transition. It describes chemical reactions in rarefied gases with external particle sources and displays a continuous non-equilibrium phase transition in the directed percolation (DP) universality class. In the mean-field (one-site) regime, the reaction channels for the first Schlögl model are written as
\begin{equation}
   A+B \xrightleftharpoons[k_2]{k_1} 2A, \quad \text{and} \quad A \xrightleftharpoons[k_4]{k_3} C,
\end{equation}
which are combined with external sources and sinks for particles $B$ and $C$ that keep their concentrations constant in time, where $\rho=N_A/N$, $b = N_B/N$, $c = N_C/N$, and $N$ is the system size ($b$ and $c$ are constants). The dynamics are entirely described by the concentration $\rho$ of $A$ particles.

The concentration of $A$ particles obeys the following time-evolution equation
\begin{equation}
   \frac{d\rho}{dt} = -k_2 \rho^2 + (k_1b - k_3)\rho + k_4c,
   \label{evolution-schlogl1}
\end{equation}
where we set $k_1b=1$, $k_2=1$, and $k_3=k_2\lambda$. The system exhibits a continuous active-absorbing phase transition for $k_4=0$ as the control parameter $\lambda$ increases. Eq.~(\ref{evolution-schlogl1}), with these settings, predicts the threshold $\lambda_c=1$. For $k_4 \ne 0$, there is no transition driven by the control parameter $\lambda$.

\begin{figure}[tbp]
   \begin{center}
      \includegraphics[scale=0.285]{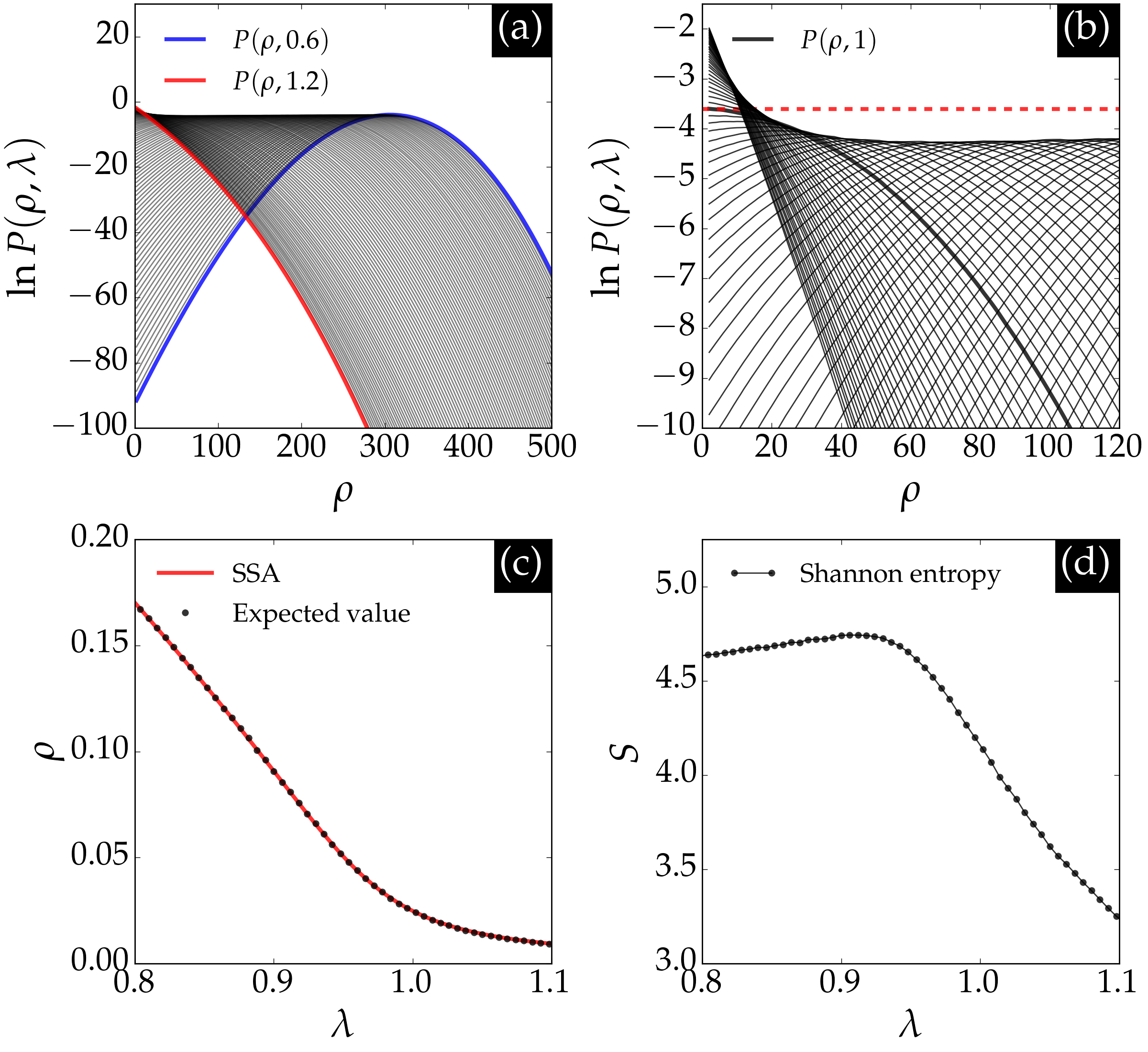}
   \end{center}
   \caption{(Color Online) Results for the first Schlögl model. As in Fig.~(\ref{results-cp}), the system is in the active phase for $\lambda<1$ and in the absorbing phase for $\lambda>1$.}
   \label{results-schlogl1}
\end{figure}

Fig.~\ref{results-schlogl1} shows the stationary distribution $P(\rho,\lambda)$ obtained using the kinetic flat-histogram algorithm. The critical distribution $P(\rho,\lambda_c)$ approaches the absorbing state with a vanishing derivative, similar to the behavior observed for the contact process. As in the active-absorbing phase transition of the contact process, the Shannon entropy remains continuous.

%%%%%%%%%%%%%%%%%%%%%%%%%%%%%%%%%%%%%%%%%%%%%%%%%%%%%%%%%%%%%%%%%%%%%%%%%%
\section{Discontinuous Phase Transitions}\label{sec-results-discontinuous}
%%%%%%%%%%%%%%%%%%%%%%%%%%%%%%%%%%%%%%%%%%%%%%%%%%%%%%%%%%%%%%%%%%%%%%%%%%

In this section, we present our findings for well-known bistable non-equilibrium stochastic systems in the mean-field (one-site) regime. Two of the most studied examples are the second Schlögl model~\cite{Schlogl-1972, Grassberger-1981, Grassberger-1982, Wang-2015} and the Ziff-Gulari-Barshad (ZGB) model~\cite{Ziff-1986, Voigt-1997, Oliveira-2016, Fernandes-2018, Hinrichsen-2000, Vilela-2020}. These two stochastic processes provide a test to determine whether kinetic flat-histogram simulations can capture the physics of discontinuous phase transitions, especially weak ones. For all results, we set the system size to $N=10^3$.

%~~~~~~~~~~~~~~~~~~~~~~~~~~~~~~~~
\subsection{Second Schlögl Model}
%~~~~~~~~~~~~~~~~~~~~~~~~~~~~~~~~

In parallel with the first Schlögl model, the second Schlögl model~\cite{Schlogl-1972, Grassberger-1981, Grassberger-1982, Wang-2015} was introduced as a prototype of an open chemical reactive system exhibiting discontinuous phase transitions. It also describes chemical reactions in rarefied gases with external particle sources. In the mean-field (one-site) regime, the reaction channels for the second Schlögl model are given by
\begin{equation}
   2A+B \xrightleftharpoons[k_2]{k_1} 3A, \quad \text{and} \quad A \xrightleftharpoons[k_4]{k_3} C.
\end{equation}
As in the first Schlögl model, particles $B$ and $C$ can be added or removed to keep their concentrations constant, where $\rho=N_A/N$, $b = N_B/N$, $c = N_C/N$, with $b$ and $c$ constants and $N$ the system size. Once again, the dynamics are entirely described by the concentration $\rho$ of $A$ particles.

The concentration of $A$ particles obeys the following time-evolution equation
\begin{equation}
   \frac{d\rho}{dt} = -k_2 \rho^3 + k_1b\rho^2 - k_3\rho + k_4c,
   \label{evolution-schlogl2}
\end{equation}
where we set $k_1b=1$, $k_2=1$, $k_3=k_2\lambda$, and $k_4c=\alpha$. Eq.~(\ref{evolution-schlogl2}), with these settings, predicts a line of discontinuous phase transitions with thresholds $\lambda_\alpha$. The critical line ends at a continuous phase transition with critical threshold $\lambda_c$ as $\alpha$ increases. When $\alpha=0$, we obtain the exact threshold of a discontinuous active-absorbing phase transition at $\lambda_0 = 1/4$, where the system is active for $\lambda<\lambda_0$ and absorbing otherwise. With our settings, the \textit{terminus} of the critical line is the critical point at $\alpha_c = 1/27$ and $\lambda_c = 1/3$~\cite{Tome-2015}.

\begin{figure}[tbp]
   \begin{center}
      \includegraphics[scale=0.285]{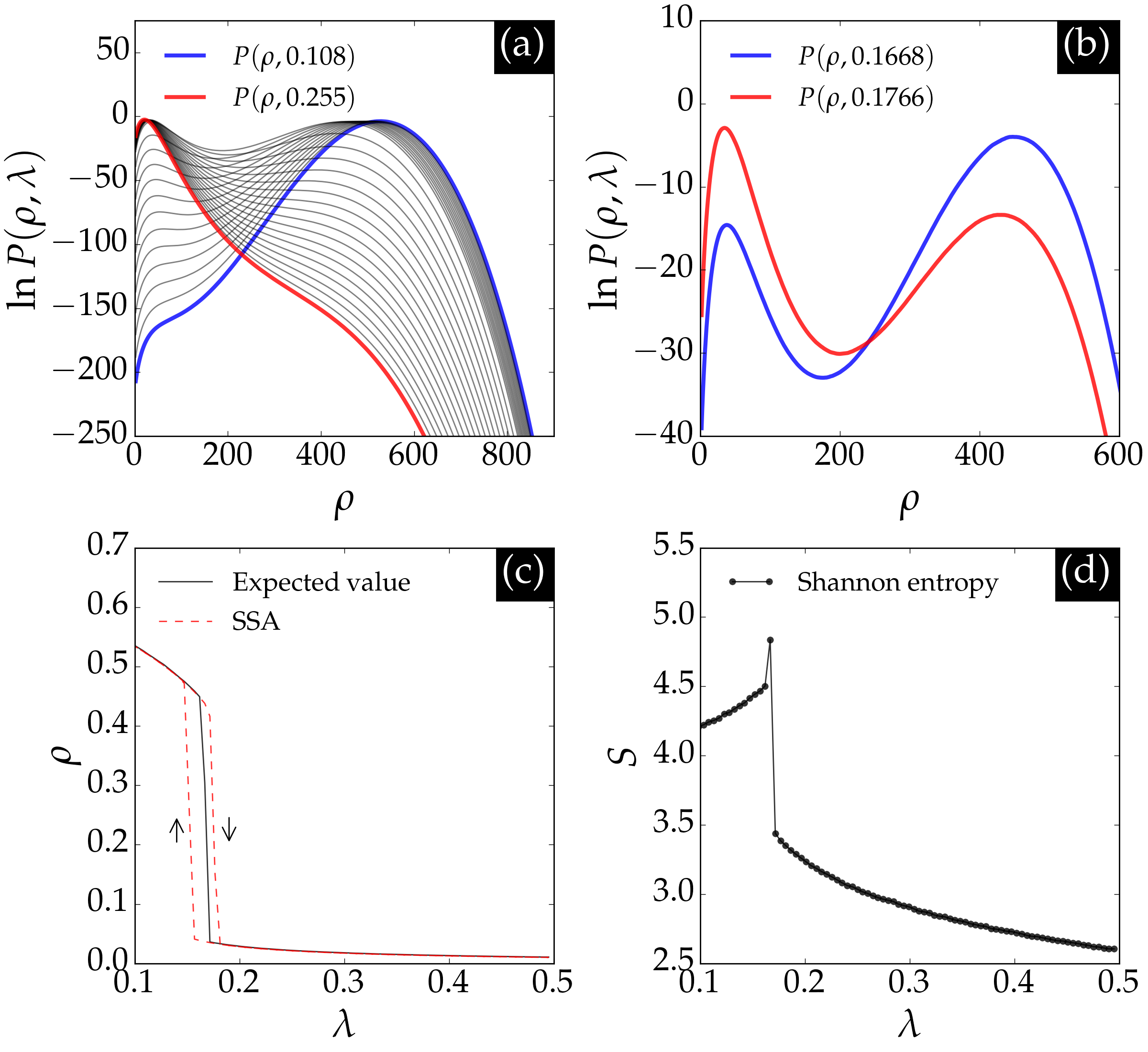}
   \end{center}
   \caption{(Color Online) Results for the second Schlögl model with $\alpha=0.005$. Panel (a) shows the stationary distribution $P(\rho,\lambda)$ for several values of the control parameter $\lambda$. We observe that $P(\rho,\lambda)$ exhibits two well-defined maxima as the threshold is approached. Panel (b) presents $P(\rho,\lambda)$ for two values of $\lambda$ close to the threshold $\lambda_\alpha$, one for each phase. Phase coexistence occurs when the two maxima of $P(\rho,\lambda)$ have equal weight. Panel (c) compares the expected value of the concentration $\rho$ with SSA simulations performed by increasing and decreasing $\lambda$. The SSA curves form a hysteresis cycle, which is a distinctive feature of discontinuous phase transitions. Finally, panel (d) shows the Shannon entropy $S(\lambda)$, which displays a clear discontinuity. The line in panel (d) is shown only as a guide to the eye.}
   \label{results-schlogl2-0.005}
\end{figure}

\begin{figure}[tbp]
   \begin{center}
      \includegraphics[scale=0.285]{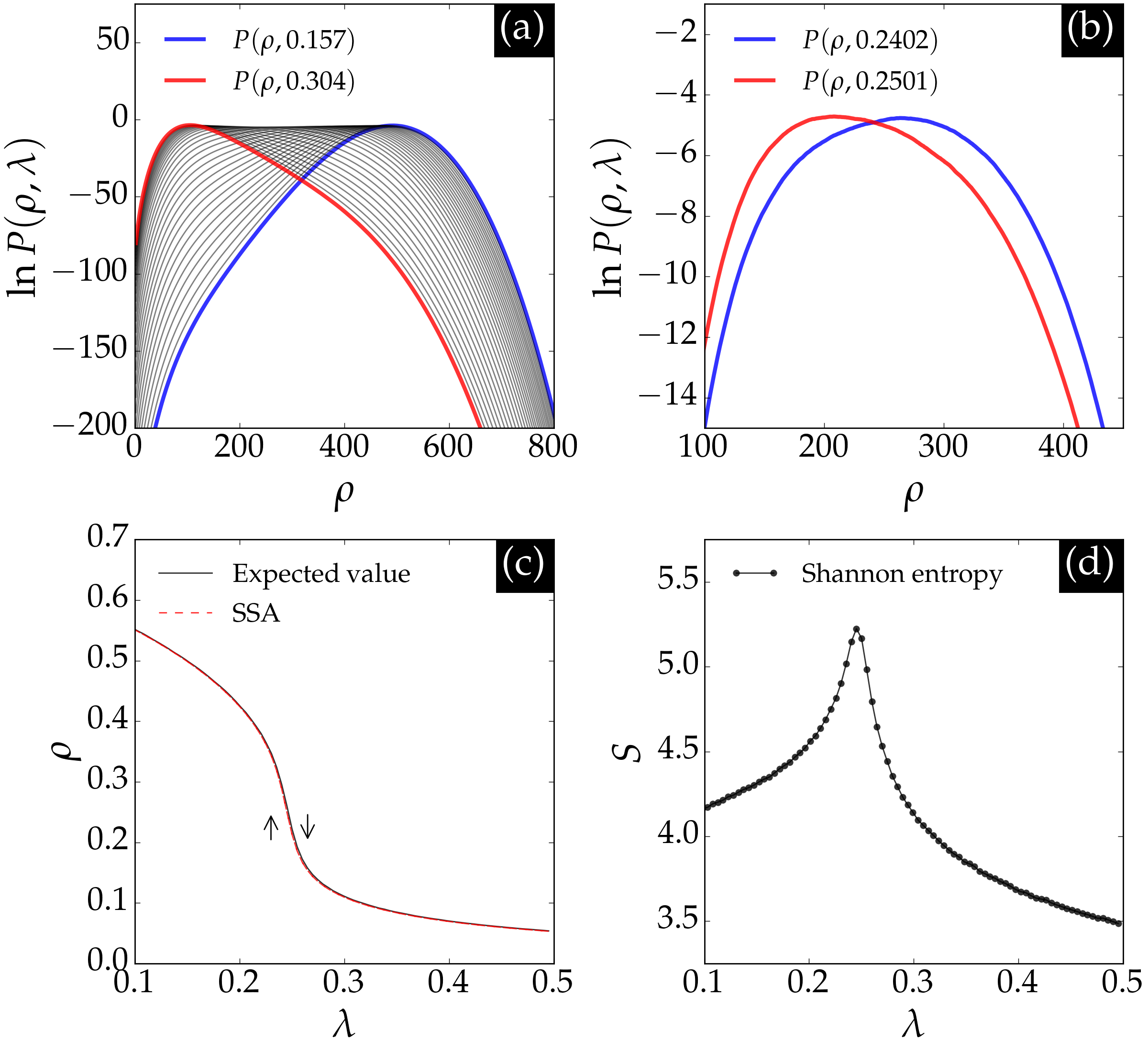}
   \end{center}
   \caption{(Color Online) Results for the second Schlögl model with $\alpha=0.025$. As in Fig.~(\ref{results-schlogl2-0.005}), we are unable to observe two distinct maxima in the stationary distribution $P(\rho,\lambda)$. However, $\ln P(\rho,\lambda)$ is clearly asymmetric, indicating a deviation from the symmetric parabolic shape. This provides evidence of a weak discontinuous phase transition, where the entropy jump is barely detectable. Additionally, the transition is too weak to exhibit a hysteresis cycle, and the Shannon entropy is practically continuous.}
   \label{results-schlogl2-0.025}
\end{figure}

\begin{figure}[tbp]
   \begin{center}
      \includegraphics[scale=0.285]{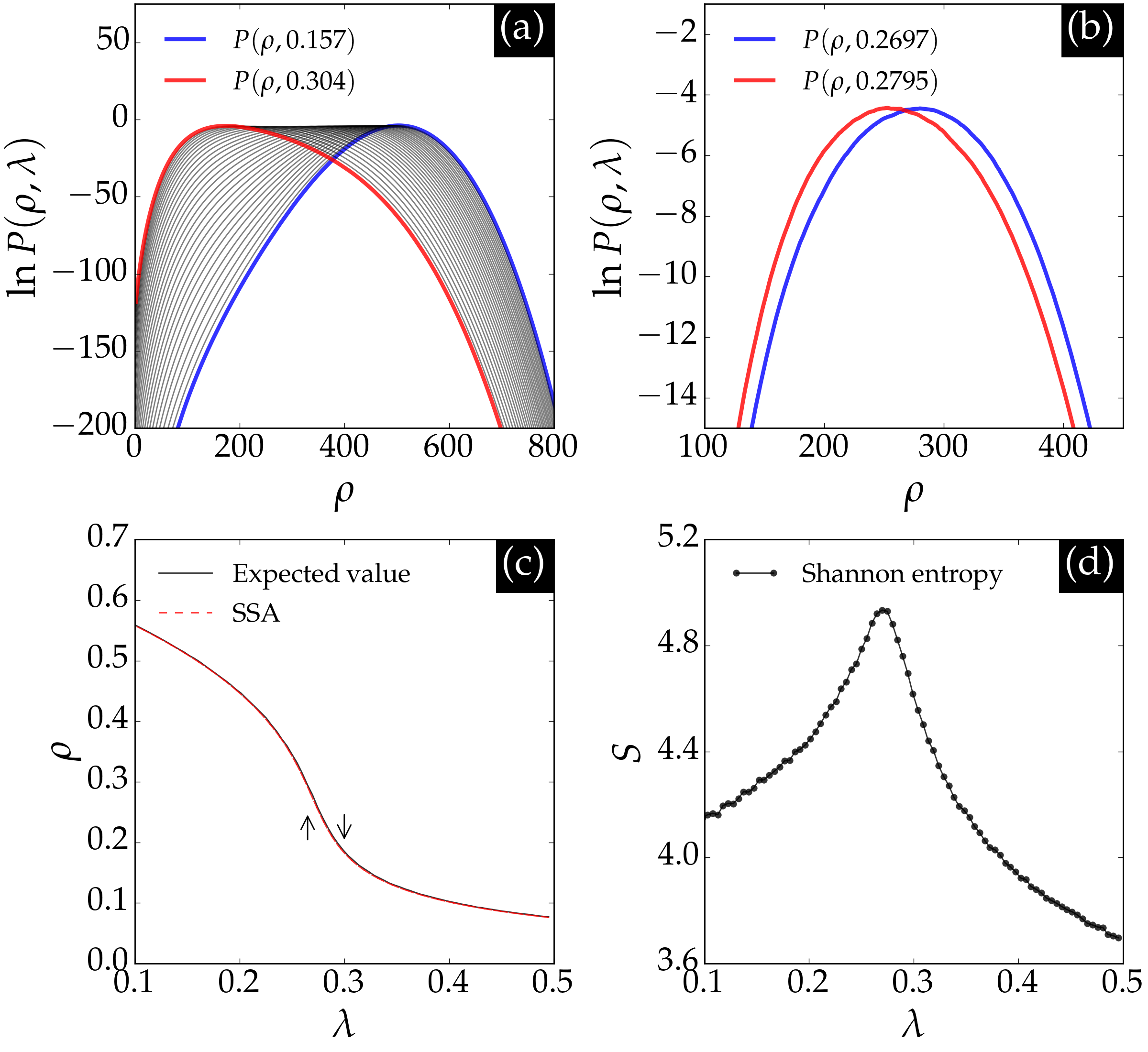}
   \end{center}
   \caption{(Color Online) Results for the second Schlögl model with $\alpha=0.035$, close to the critical point at $\alpha_c=1/27 \sim 0.037$ and $\lambda_c = 1/3 \sim 0.333$. As in Fig.~(\ref{results-schlogl2-0.005}), two distinct maxima are no longer visible in the stationary distribution $P(\rho,\lambda)$. However, $\ln P(\rho,\lambda)$ is nearly symmetric and close to a parabolic shape, indicating proximity to the critical point. Consequently, the transition is too weak to exhibit any hysteresis cycle, and the Shannon entropy is practically continuous.}
   \label{results-schlogl2-0.035}
\end{figure}

We present results for the second Schlögl model in Figs.~(\ref{results-schlogl2-0.005}), (\ref{results-schlogl2-0.025}), and (\ref{results-schlogl2-0.035}), where $\alpha$ is increased as the \textit{terminus} of the critical line is approached. For a strong discontinuous phase transition, as shown in Fig.~(\ref{results-schlogl2-0.005}), all the distinctive features of a discontinuous phase transition are evident. Even the SSA simulations reveal the hysteresis cycle. However, the hysteresis cycle is sensitive to the number of Monte Carlo steps (or the simulation time) discarded to allow the system to reach the stationary distribution. The stationary distribution displays two clear maxima, and the threshold can be estimated by identifying the $P(\rho,\lambda)$ corresponding to phase coexistence, when the two maxima have equal weight. Additionally, the Shannon entropy exhibits a clear jump near the threshold.

Fig.~(\ref{results-schlogl2-0.025}) presents results for a weak discontinuous phase transition. It becomes challenging to identify a discontinuous phase transition using standard methods. We discarded a sufficiently long initial simulation time to allow the system to reach the stationary distribution, so no hysteresis cycle is observed. Moreover, the jump in the Shannon entropy at the threshold begins to vanish. To observe clearer effects, larger system sizes should be used. Nevertheless, a distinctive feature remains: $\ln P(\rho,\lambda)$ is clearly asymmetric, allowing us to identify a possible weak discontinuous phase transition.

Fig.~\ref{results-schlogl2-0.035} presents results near the critical point. It is difficult to observe any hysteresis cycle because importance sampling tunnels easily through phase space. Additionally, the Shannon entropy remains continuous. The only evidence of a discontinuous phase transition is the slight asymmetry of $\ln P(\rho,\lambda)$ near the maxima, indicating a minor deviation from the parabolic shape. If $\alpha$ is increased further, $\ln P(\rho,\lambda)$ near the maxima becomes parabolic and symmetric.

%~~~~~~~~~~~~~~~~~~~~~~~~~~~~~~~~~~~~~
\subsection{Ziff-Gulari-Barshad Model}
%~~~~~~~~~~~~~~~~~~~~~~~~~~~~~~~~~~~~~

Another convenient test for the kinetic flat-histogram algorithm is the ZGB model~\cite{Ziff-1986, Voigt-1997, Oliveira-2016, Fernandes-2018, Hinrichsen-2000, Vilela-2020}, which describes the catalytic oxidation of a metallic surface via the Langmuir-Hinshelwood process. The Langmuir-Hinshelwood process consists of three steps for a chemical reaction: (1) adsorption of the reactants onto a catalytic surface, (2) recombination of the reactants, and (3) desorption of the products.

The reaction channels of the ZGB model are given by
\begin{eqnarray}
   && C \xrightarrow{k_1} A, \quad 2C \xrightarrow{k_2} 2B, \nonumber \\
   && A+B \xrightarrow{k_3} 2C, \quad \text{and} \quad A \xrightarrow{k_4} C,
   \label{channels-zgb}
\end{eqnarray}
where $A$, $B$, and $C$ represent a carbon monoxide molecule, an oxygen atom, and an empty surface site, respectively. The channels from left to right in Eq.~(\ref{channels-zgb}) are interpreted as follows:
\begin{itemize}
   \item First channel: adsorption of a carbon monoxide molecule, which occupies an empty site;
   \item Second channel: adsorption of an oxygen molecule, where each of the two resulting atoms occupies an empty site;
   \item Third channel: oxidation of carbon monoxide, resulting in a carbon dioxide molecule that is ejected from the surface, leaving two empty sites;
   \item Last channel: possibility of a carbon monoxide molecule being ejected from the surface without recombination, leaving a single empty site. 
\end{itemize}
This last channel allows for the existence of a critical line with a \textit{terminus} at a critical point, as observed in the second Schlögl model.

Let $x_1$, $x_2$, and $x_3$ denote the concentrations of particles $A$, $B$, and $C$, respectively. The coupled time-evolution equations for the concentrations are given by
\begin{equation}
   \left\lbrace
   \begin{aligned}
      &\frac{dx_1}{dt} = k_1 x_3 - k_4 x_1 - k_3 x_1 x_2, \\
      &\frac{dx_2}{dt} = 2 k_2 x_3^2 - k_3 x_1 x_2, \\
      &\frac{dx_3}{dt} = -k_1 x_3 + k_4 x_1 - 2 k_2 x_3^2 + 2 k_3 x_1 x_2. \\
   \end{aligned}
   \right.
   \label{evolution-zgb}
\end{equation}
Only two of the three equations are independent. The three concentrations are bounded, and the dynamics can be described by analyzing the carbon monoxide concentration, where we define $\rho \equiv x_1$. Furthermore, we set $k_1 = \lambda$, $k_2 = 0.5$, $k_3=1$, and $k_4=\alpha$. Eq.~(\ref{evolution-zgb}) predicts a line of discontinuous phase transitions with thresholds $\lambda_\alpha$ ending at a critical point. With these settings, the \textit{terminus} of the critical line can be calculated numerically, yielding $\alpha_c \sim 0.08$~\cite{Tome-2015}.

\begin{figure}[tbp]
   \begin{center}
      \includegraphics[scale=0.285]{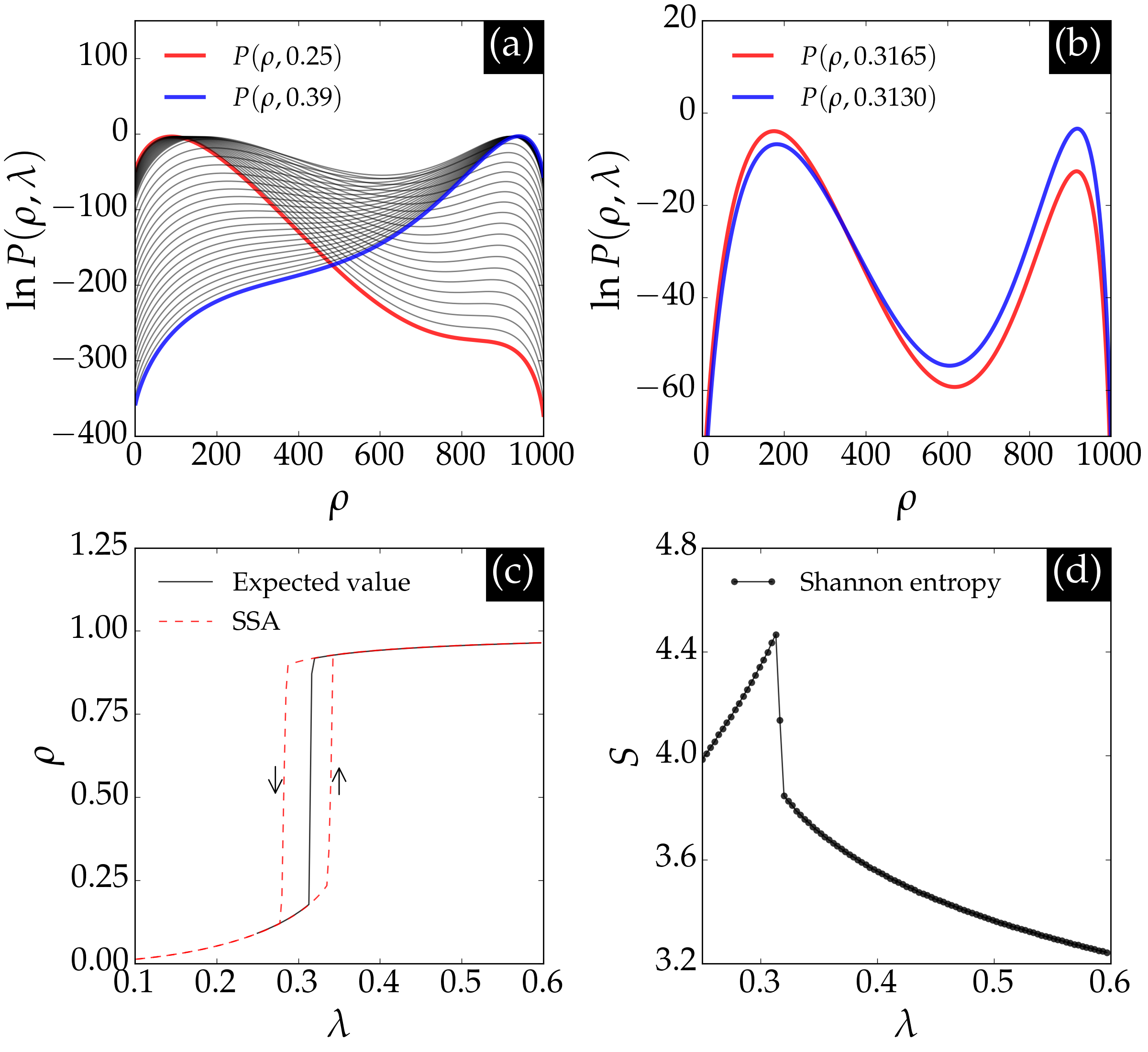}
   \end{center}
   \caption{(Color Online) Results for the ZGB model with $\alpha=0.02$. Panel (a) shows the stationary distribution $P(\rho,\lambda)$ of the carbon monoxide concentration $\rho$ adsorbed on the surface for several values of the control parameter $\lambda$. We observe that $P(\rho,\lambda)$ exhibits two well-defined maxima as the threshold is approached. Panel (b) presents $P(\rho,\lambda)$ for two values of $\lambda$ close to the threshold $\lambda_\alpha$, one for each phase. Phase coexistence occurs when the two maxima of $P(\rho,\lambda)$ have equal weight. Panel (c) compares the expected value of the concentration $\rho$ with SSA simulations performed by increasing and decreasing $\lambda$. The SSA curves form a hysteresis cycle, which is a distinctive feature of discontinuous phase transitions. Finally, panel (d) shows the Shannon entropy $S(\lambda)$, which is clearly discontinuous. The line in panel (d) is shown only as a guide to the eye.}
   \label{results-zgb-0.02}
\end{figure}

\begin{figure}[tbp]
   \begin{center}
      \includegraphics[scale=0.285]{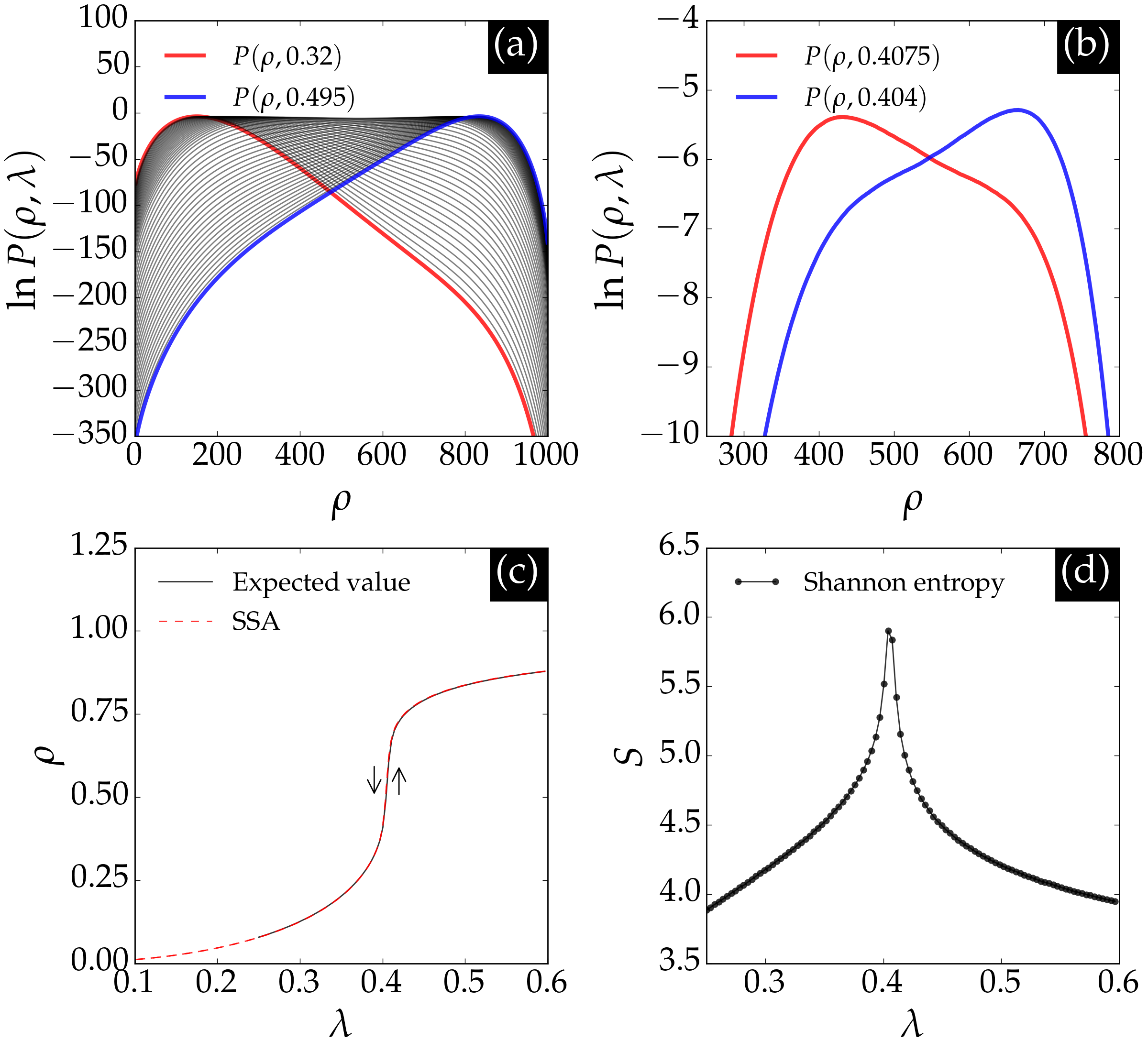}
   \end{center}
   \caption{(Color Online) Results for the ZGB model with $\alpha=0.06$. As in Fig.~(\ref{results-zgb-0.02}), it becomes difficult to observe two distinct maxima in the stationary distribution $P(\rho,\lambda)$. However, $\ln P(\rho,\lambda)$ is clearly asymmetric, indicating a deviation from the symmetric parabolic shape. This provides evidence of a weak discontinuous phase transition, where the entropy jump is barely detectable. Additionally, the transition is too weak to exhibit a hysteresis cycle, and the Shannon entropy is practically continuous.}
   \label{results-zgb-0.06}
\end{figure}

\begin{figure}[tbp]
   \begin{center}
      \includegraphics[scale=0.285]{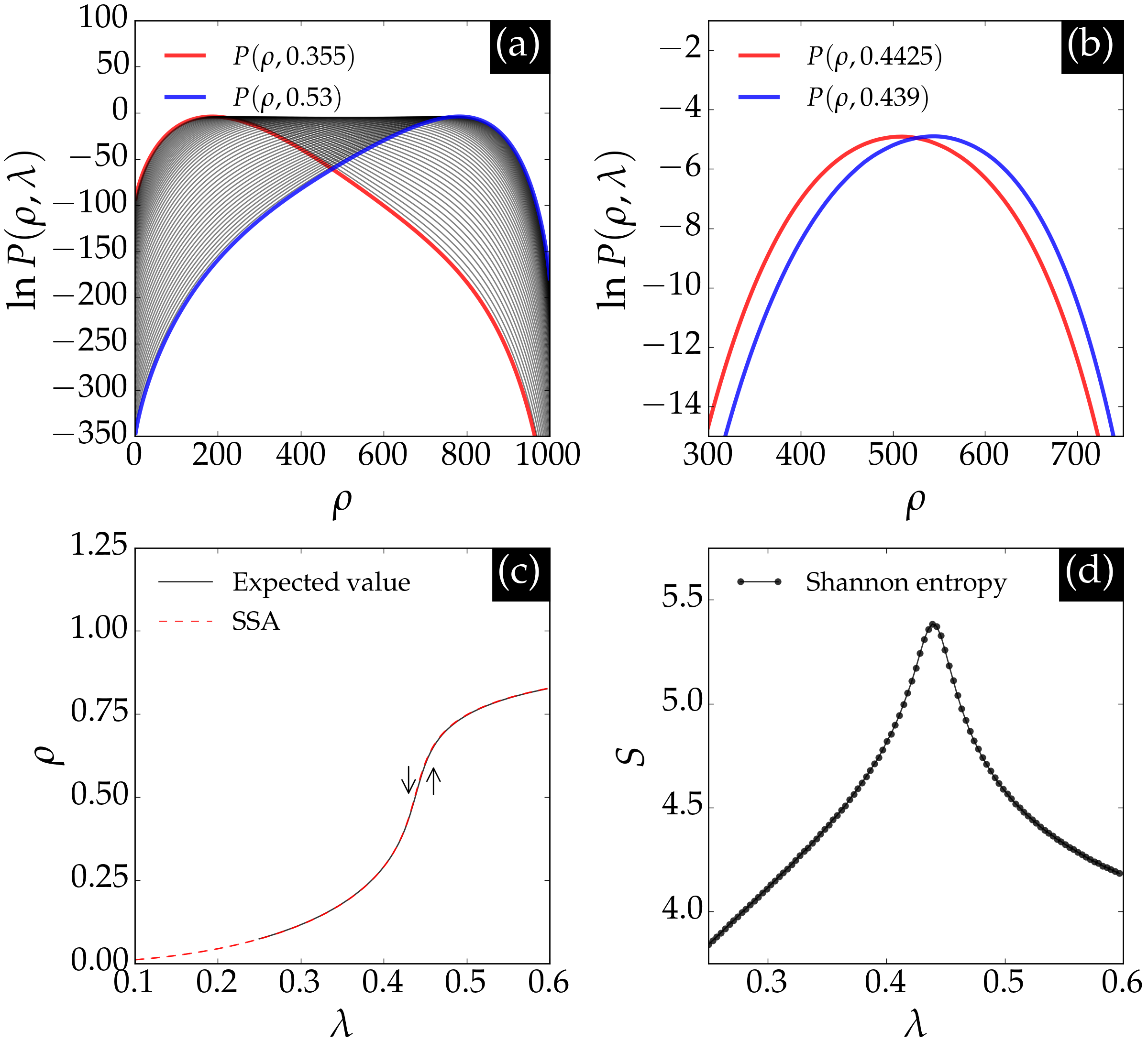}
   \end{center}
   \caption{(Color Online) Results for the ZGB model at the critical point $\alpha_c=0.08$. As in Fig.~(\ref{results-zgb-0.02}), $\ln P(\rho,\lambda)$ is now nearly symmetric and very close to a parabolic shape. Consequently, there is no hysteresis cycle, and the Shannon entropy is practically continuous.}
   \label{results-zgb-0.08}
\end{figure}

We present results for the ZGB model in Figs.~(\ref{results-zgb-0.02}), (\ref{results-zgb-0.06}), and (\ref{results-zgb-0.08}). The behavior of the stationary distribution $P(\rho,\lambda)$ is similar to that observed for the second Schlögl model as the critical point is approached. Far from the critical point, as in Fig.~(\ref{results-zgb-0.02}), we observe a strong discontinuous phase transition, where hysteresis cycles can be easily identified using standard methods. The Shannon entropy displays a clear step, and the threshold can be estimated by identifying the $P(\rho,\lambda)$ with two maxima of equal weight.

As the critical point is approached by increasing $\alpha$, the hysteresis cycle eventually disappears. The step in the Shannon entropy also vanishes. Nevertheless, the asymmetry of $\ln P(\rho,\lambda)$ persists even near the critical point, as seen in Fig.~(\ref{results-zgb-0.06}). In Fig.~(\ref{results-zgb-0.08}), we show the behavior of $\ln P(\rho,\lambda)$ close to the critical point, where it becomes symmetric and nearly parabolic.

%%%%%%%%%%%%%%%%%%%%%%%%%%%%%%%%%%%%%%%%%%%%
\section{Conclusions}\label{sec-conclusions}
%%%%%%%%%%%%%%%%%%%%%%%%%%%%%%%%%%%%%%%%%%%%

We introduced the kinetic flat-histogram algorithm, which is effective for identifying weak discontinuous phase transitions in ergodic stochastic processes. The algorithm performs a random walk in phase space to estimate the stationary distribution of an observable in a non-equilibrium stochastic process. The kinetic flat-histogram algorithm estimates the stationary distribution of an observable by rejection sampling, generating trial moves using standard Monte Carlo techniques with an acceptance probability inversely related to the estimated stationary distribution.

The stationary distribution is refined during the random walk through a modification factor, which introduces non-Markovian dynamics. The process becomes Markovian as the estimated stationary distribution converges to its asymptotic form. A visitation histogram is accumulated, and the modification factor is updated when this histogram satisfies a flatness condition. The simulation ends when the modification factor reaches a threshold value close to unity.

The update procedure assigns greater weights to the most frequently visited states. Since the acceptance probability is inversely related to the stationary distribution, the algorithm encourages visits to less degenerate states (rare events). Thus, the flat-histogram condition acts as a balancing mechanism that equalizes the visitation of different macroscopic states.

Similar to the Wang-Landau algorithm, the kinetic flat-histogram algorithm is also prone to error saturation. However, this issue can be mitigated by employing strategies such as the $1/t$ method or adaptive refinement schemes. A theoretical analysis providing a definitive proof of convergence when using $1/t$ updates is still lacking and remains an open question for future research.

%%%%%%%%%%%%%%%%%%%%%%%%%%%%%%%%%%%%%%%%%%%%%%%%%%%%%%
\section{Acknowledgments} \label{sec:acknowledgements}
%%%%%%%%%%%%%%%%%%%%%%%%%%%%%%%%%%%%%%%%%%%%%%%%%%%%%%

We gratefully acknowledge financial support from CAPES (Coordenação de Aperfeiçoamento de Pessoal de Nível Superior), CNPq (Conselho Nacional de Desenvolvimento Científico e Tecnológico), and FAPEPI (Fundação de Amparo à Pesquisa do Estado do Piauí). We thank the \textit{Dietrich Stauffer Computational Physics Lab}, Teresina, Brazil, and the \textit{Laborat\'{o}rio de F\'{\i}sica Te\'{o}rica e Modelagem Computacional (LFTMC)}, Teresina, Brazil, for providing computational resources for the numerical simulations. R. S. Ferreira acknowledges support from FAPEMIG (grant FAPEMIG-APQ-06611-24).

\newpage

\bibliography{textv4}

\end{document}